\documentclass[journal]{IEEEtran}  % For two columns
\ifCLASSINFOpdf
\else
\fi
\usepackage{amsmath,graphicx,cite}
\usepackage{subfig}
\usepackage{widetext}
\usepackage{algorithm}
\usepackage{algcompatible}
\usepackage[end]{algpseudocode}

\makeatletter
\def\BState{\State\hskip-\ALG@thistlm}
\makeatother

\newcommand{\myindent}[1]{
\newline\makebox[#1cm]{}
}

\begin{document}

%\title{N-type PHD Filter for Multiple Target, Multiple Type Filtering}
\title{Multiple Target, Multiple Type Filtering in the RFS Framework}
%\title{Extended Probability Hypothesis Density Filter for Multi-class Multi-target tracking}

\author{Nathanael~L.~Baisa,~\IEEEmembership{Member,~IEEE,}
        Andrew~Wallace,~\IEEEmembership{Fellow,~IET}
        %and~Daniel~Clark,~\IEEEmembership{Fellow,~IEEE}% <-this % stops a space
       % \\
      %Department of Electrical, Electronic and Computer Engineering \\
      %School of Engineering and Physical Sciences, \\
      %Heriot Watt University, Edinburgh, United Kingdom, \\
      %E-mail: \{nb30, a.m.wallace\}@hw.ac.uk %\vspace{-5mm}
      %E-mail: \{nb30, a.m.wallace, d.e.clark\}@hw.ac.uk %\vspace{-5mm}
\thanks{N. L. Baisa and A. Wallace are with the Department of Electrical, Electronic and Computer Engineering, Heriot Watt University, Edinburgh EH14 4AS, United Kingdom. (e-mail: \{nb30, a.m.wallace\}@hw.ac.uk).}}% <-this % stops a space
\maketitle

\begin{abstract}
%We develop a new filter based on the standard Probability Hypothesis Density (PHD) filter for filtering multiple type of different targets for handling confusions between them. In this approach, we assume there will be confusions between detections, i.e. clutter arises not just from background false positives, but also from target confusions. Under the Gaussianity and linearity assumptions, we also propose its Gaussian mixture implementation (N-type GM-PHD filter) which shows improved performance over the standard approach.

A Multiple Target, Multiple Type Filtering (MTMTF) algorithm is developed using Random Finite Set (RFS) theory. First, we extend the standard Probability Hypothesis Density (PHD) filter for multiple types of targets, each with distinct detection properties, to develop a multiple target, multiple type filtering, N-type PHD filter, where $N\geq2$, for handling confusions among target types. In this approach, we assume that there will be confusions between detections, i.e. clutter arises not just from background false positives, but also from target confusions. Then, under the assumptions of Gaussianity and linearity, we extend the Gaussian mixture (GM) implementation of the standard PHD filter for the proposed N-type PHD filter termed the N-type GM-PHD filter. Furthermore, we analyze the results from simulations to track sixteen targets of four different types using a four-type (quad) GM-PHD filter as a typical example and compare it with four independent GM-PHD filters using the Optimal Subpattern Assignment (OSPA) metric. This shows the improved performance of our strategy that accounts for target confusions by efficiently discriminating them\footnote{https://github.com/nathanlem1/MTF-Lib}.

\end{abstract}

\begin{IEEEkeywords}
Random finite set, FISST, Multiple target filtering, PHD filter, N-type PHD filter, Gaussian mixture, OSPA metric
\end{IEEEkeywords}

% \fi
\IEEEpeerreviewmaketitle

\section{Introduction}

Multi-target filtering is a state estimation problem which plays a key role in visual, radar and sonar tracking, robot simultaneous localization and mapping (SLAM), and other signal processing applications. Traditionally, multi-target filters are based on finding associations between targets and measurements using methods including Global Nearest Neighbour (GNN)~\cite{BarFor88}\cite{CaiFreLit06}, Joint Probabilistic Data Association Filter (JPDAF)~\cite{BarFor88}\cite{RasHag01}, and Multiple Hypothesis Tracking (MHT)~\cite{BarFor88}\cite{ChaReh99}. However, these approaches face challenges not only in the uncertainty caused by the data association but also in computational growth exponential to the number of targets and measurements. To address the complexity problem, a unified framework directly extended single- to multi-target tracking by representing multi-target states and observations as a random finite set (RFS)~\cite{Mah03}. This estimates both the states and cardinality of an unknown and time varying number of targets in a scene, and includes birth, death, clutter (false alarms), and missed detections. Mahler~\cite{Mah03} propagated the first-order moment of the multi-target posterior called Probability Hypothesis Density (PHD) or intensity rather than the full multi-target posterior, thus having much lower computational complexity in the single state space rather than in the joint-state space as in traditional methods. Further developments include the Gaussian mixture (GM-PHD)~\cite{VoMa06} and the Sequential Monte Carlo (SMC) (particle-PHD filter) implementations~\cite{VoSinDou05} which have been applied to visual tracking in~\cite{ZhoLi14} and~\cite{MagTajCav08}, respectively. This approach is flexible, for instance, it has been used to find the detection proposal with the maximum weight as the target position estimate for tracking a target of interest in dense environments by removing the other detection proposals as clutter~\cite{BaiBhoWal17}\cite{BaiBhoWal18}. Furthermore, the PHD filter has also been used for doing visual odometry (VO)~\cite{ZhaStaGas12} and SLAM~\cite{AdaVoMah14} in robotics. Joint detection, tracking and classification (JDTC) of multiple targets in clutter which jointly estimates the number of targets, their kinematic states, and types of targets (classes) from a sequence of noisy and cluttered observation sets was developed using the PHD filter in~\cite{YanFuLonLi12}. In this approach, the dynamics of each target type (class) is modelled as a class-dependent model set, and the signal amplitude is included within the multi-target likelihood in the PHD-like filter to enhance the discrimination between targets from different classes and false alarms. Similarly, a joint target tracking and classification (JTC) algorithm was developed in~\cite{YanFuLi14} using RFS which takes into account extraneous target-originated measurements (of the same type) i.e. multiple measurements originating from a target which can be modeled as a Poisson RFS. In these approaches, the augmented state vector of a target comprises the target kinematic state and class label, i.e. the target type (class) is put into the target state vector. However, all these RFS-based multiple target filters were developed for either a single target type or multiple target type but without taking any account of target confusions between target types at the measurement stage, i.e. measurements originated not only from the same target type but were also confused from the other target types.

%Practically, in some cases, for example for situational awareness, driver assistance and vehicle autonomy, there is also a necessity to distinguish between different target types, between vehicles and more vulnerable road users such as pedestrians and bicycles to select the best sensor focus and course of action~\cite{Matzka2012}.
%For sports analysis we often want to track and discriminate sub-groups of the same target type such as players in opposing teams~\cite{LiuCarr14}. In this and many other examples, confusion between target types is common; a standard histogram-based detection strategy~\cite{Dollar2012} in an urban environment may provide confused detections between pedestrians and cyclists, and even small cars.

Practically, there are many situations where tracking and discrimination of multiple target types is essential, handling confusions between target types. For example, when developing situational awareness for driver assistance and vehicle autonomy~\cite{Matzka2012}, a vehicle equipped with a sensor suite must detect and track other road users to select the best sensor focus and course of action, usually concentrating on other vehicles, pedestrians and bicycles. In this particular and many other examples, confusion between target types is common, for example a standard pedestrian detection strategy~\cite{DolAppPerBel14} often provides confused detections between pedestrians and cyclists, and even small cars. Moreover, for sports analysis we often want to track and discriminate sub-groups of the same target type such as players in opposing teams~\cite{LiuCarr14}. These types of problems motivate our work to develop a multiple target, multiple type filtering methodology handling target confusions following the RFS based filtering method, particularly the standard PHD filter, without involving any data association.

%In this paper, we make the following contributions. First, we model filtering of N-types of multiple targets with separate but confused detections or measurements based on standard PHD filter which we call N-type PHD filter where $N\geq2$ described in sections~\ref{Sec:FilteringWithRFS}, ~\ref{Subsec:PGFL} and~\ref{Subsec:DualPHDfilter}. Second, Gaussian mixture implementation of the standard PHD filter is extended for the proposed N-type PHD filter which is termed as N-type GM-PHD filter in section~\ref{Subsec:GM-DualPHD}. We demonstrate this proposed N-type GM-PHD filter by simulations, specifically for a quad GM-PHD filter ($N = 4$) as a typical example in section~\ref{Sec:ExperimentalResults}. The main conclusions and suggestions for future work are summarized in section~\ref{Sec:Conclusion}.

The main contributions of this paper are as follows.
\begin{itemize}
  \item We model the RFS filtering of $N$ different types of multiple targets with separate but confused detections which we call the N-type PHD filter where $N\geq2$.
  \item The Gaussian mixture implementation of the standard PHD filter is extended for the proposed N-type PHD filter.
  \item We demonstrate this proposed N-type GM-PHD filter by simulations, specifically for a quad GM-PHD filter ($N = 4$) as a typical example under different values of confusion detection probabilities to show that our approach yields improved performance over the standard approach.
\end{itemize}

We presented preliminary ideas in~\cite{BaiWal17} for three different target types (N=3) for visual tracking applications. In this work, we further develop our approach. We extend from a tri-PHD filter to a N-type PHD filter as well as conducting experiments on more dense simulations. The mathematical proof of the algorithm is presented. We demonstrate the quad GM-PHD filter for four types (N = 4) of sixteen targets with detailed analysis as a typical simulation example under different values of confusion detection probabilities.

The remainder of this paper is organized as follows. Multiple type, multiple target recursive Bayes filtering with RFS is described in section~\ref{Sec:FilteringWithRFS}. A probability generating functional for deriving a N-type PHD filter and the N-type PHD filtering strategy are given in sections~\ref{Subsec:PGFL} and~\ref{Subsec:DualPHDfilter}, respectively. In section~\ref{Subsec:GM-DualPHD}, a Gaussian mixture implementation of the N-type PHD filter is described in detail. The experimental results are analyzed and compared in section~\ref{Sec:ExperimentalResults}. The main conclusions and suggestions for future work are summarized in section~\ref{Sec:Conclusion}.

%\hfill September 17, 2014

\section{Multiple Target, Multiple Type Recursive Bayes Filtering with RFS} \label{Sec:FilteringWithRFS}

A RFS represents a varying number of non-ordered target states and observations which is analogous to a random vector for single target tracking.
More precisely, a RFS is a finite-set-valued random variable i.e. a random variable which is random in both the number of elements and the values of the elements themselves.
Finite Set Statistics (FISST), the study of statistical properties of RFS, is a systematic treatment of multi-sensor multi-target filtering as a unified Bayesian framework using random set theory~\cite{Mah03}.

%When many different detectors run on the same scene to detect many target types such as pedestrians, cars, vans, each football team, etc, there is no guarantee that these detectors only detect their own type i.e. the detector of one target type may detect the other target types and vice versa. If we assume all detectors only detect their own type, independent PHD filter for each target type can be used to filter them independently. In this paper, we assume that one target type detector is more likely to give a positive response where the other target types are present than only from the background, and vice versa.

When different detectors run on the same scene to detect different target types, there is no guarantee that these detectors only detect their own type. It is possible to run an independent PHD filter for each target type, but this will not be correct in most cases, as the likelihood of a positive response to a target of the wrong type will in general be different from, usually higher than, the likelihood of a positive response to the scene background. In this paper, we account for this difference between background clutter and target type confusion. This is equivalent to a single sensor (e.g. a smart camera) that has N different detection modes, each with its own probability of detection and a measurement density for N different target types.

So we model a N-type PHD filter to filter N-types of multiple targets in such a way that the first PHD update will filter the first target type treating the others as potential, additional clutter in addition to background clutter, and vice versa. In the joint tracking and classification approaches such as in~\cite{YanFuLonLi12},\cite{YanFuLi14}, the target type (class) is put into the target state vector, however, here we follow a different approach which is convenient for handling target confusions from different target types. Accordingly, to derive the N-type PHD filter, it is necessary to first give its RFS representation to extend from a single type single-target Bayes framework to multiple type multi-target Bayes framework. Let the multi-target state space $\mathcal{F}(\mathcal{X})$ and the multi-target observation space $\mathcal{F}(\mathcal{Z})$ be the respective collections of all the finite subsets of the state space $\mathcal{X}$ and observation space $\mathcal{Z}$, respectively. If $L_i(k)$ is the number of targets of target type $i$ in the scene at time $k$, then the multiple states for target type $i$, $X_{i,k}$, is the set

\begin{equation}
    X_{i,k} =  \{x_{i,k,1},...x_{i,k,L_i(k)}\} \in \mathcal{F}(\mathcal{X})
\label{eqn:stateSet1}
\end{equation}
%\noindent
%\begin{equation}
%    X_{2,k} =  \{x_{k,1},...x_{k,M_2(k)}\} \in \mathcal{F}(\mathcal{X})
%\label{eqn:stateSet2}
%\end{equation}
\noindent where $i \in \{1, ..., N\}$. Similarly, if $M_j(k)$ is the number of received observations from detector $j$, then the corresponding multiple target measurements is the set

\begin{equation}
    Z_{j,k} =  \{z_{j,k,1},...z_{j,k,M_j(k)}\} \in \mathcal{F}(\mathcal{Z})
\label{eqn:observationSet1}
\end{equation}
%\noindent
%\begin{equation}
%    Z_{2,k} =  \{z_{k,1},...z_{k,N_2(k)}\} \in \mathcal{F}(\mathcal{Z})
%\label{eqn:observationSet2}
%\end{equation}
\noindent where $j \in \{1, ..., N\}$. Some of these observations may be false i.e. due to clutter (background) or
confusion (response due to another target type).

The uncertainty in the state and measurement is introduced by modeling the multi-target state and the multi-target measurement using RFS. Let $\Xi_{i,k}$ be the RFS associated with the multi-target state of target type $i$, then

\begin{equation}
    \Xi_{i,k} =  S_{i,k}(X_{i,k-1}) \cup \Gamma_{i,k},
\label{eqn:stateRFS1}
\end{equation}
%\noindent
%\begin{equation}
%    \Xi_{2,k} =  S_{2,k}(X_{2,k-1}) \cup \Gamma_{2,k},
%\label{eqn:stateRFS2}
%\end{equation}
\noindent where $S_{i,k}(X_{i,k-1})$ denotes the RFS of surviving targets of target type $i$, and $\Gamma_{i,k}$ is the RFS of the new-born targets of target type $i$. We do not consider target spawning in this paper.

Further, the RFS $\Omega_{ji,k}$ associated with the multi-target measurements of target type $i$ from detector $j$ is
\begin{equation}
    \Omega_{ji,k} =  \Theta_{j,k}(X_{i,k}) \cup C_{si,k} \cup C_{tiJ,k},
\label{eqn:observationRFS1}
\end{equation}
%\noindent
%\begin{equation}
%    \Omega_{2,k} =  \Theta_{2,k}(X_{2,k}) \cup C_{s2,k} \cup C_{t21,k},
%\label{eqn:observationRFS2}
%\end{equation}
\noindent where $J=\{1, ..., N\}\setminus i$ and $\Theta_{j,k}(X_{i,k})$ is the RFS modeling the measurements generated by the target $X_{i,k}$ from detector $j$, and $C_{si,k}$ models the RFS associated with the clutter (false alarms) for target type $i$ which comes from the scene background (using detector $j$). However, we now must also include $C_{tiJ,k}$ which is the RFS associated with measurements of all target types $J=\{1, ..., N\}\setminus i$ being treated as confusion while filtering target type $i$ i.e. measurements of all target types are included into clutter (confusion) except measurement of target type $i$ while filtering target type $i$.

Analogously to the single-target case, the dynamics of $\Xi_{i,k}$ are described by the multi-target transition density $y_{i,k|k-1}(X_{i,k}|X_{i,k-1})$, while $\Omega_{ji,k}$ is described by the multi-target likelihood $f_{ji,k}(Z_{j,k}|X_{i,k})$ for multiple target type $i \in \{1, ..., N\}$ from detector $j \in \{1, ..., N\}$. The recursive equations are

%\begin{widetext}
\begin{equation}
\begin{array} {lll}
p_{i,k|k-1}(X_{i,k}|Z_{j,1:k-1}) =\\ \int y_{i,k|k-1}(X_{i,k}|X)p_{i,k-1|k-1}(X|Z_{j,1:k-1})\mu(dX)
\end{array}
\label{eqn:predictionRFS}
\end{equation}
%\end{widetext}
\noindent
%\begin{widetext}
\begin{equation}
\begin{array} {lll}
    p_{i,k|k}(X_{i,k}|Z_{j,1:k}) =  \frac {f_{ji,k}(Z_{j,k}|X_{i,k})p_{i,k|k-1}(X_{i,k}|Z_{j,1:k-1})}{\int f_{ji,k}(Z_{j,k}|X)p_{i,k|k-1}(X|Z_{j,1:k-1})\mu(dX)}
\end{array}
\label{eqn:updateRFS}
\end{equation}
%\end{widetext}
\noindent where $\mu$ is an appropriate dominating measure on $\mathcal{F}(\mathcal{X})$~\cite{Mah03}.
Though a Monte Carlo approximation of this optimal multi-target types Bayes recursion is possible considering multiple targets of a single type~\cite{VoSinDou05}, the number of particles required is exponentially related to the number of targets and their types in the scene. To make it computationally tractable, we extended Mahler's method of propagating the first-order moment of the multi-target posterior instead of the full multi-target posterior as its approximation called the Probability Hypothesis Density (PHD)~\cite{Mah03}, $\mathcal{D}_{i,k|k}(x|Z_{j,1:k}) = \mathcal{D}_{i,k|k}(x) = \int \delta_x(x) p_{i,k|k}(X|Z_{j,1:k})\delta X$ where $\delta_x(x) = \sum_{w \in x}\delta_w(x)$, for $N\geq2$ types of multiple targets by deriving the updated PHDs from Probability Generating Functionals (PGFLs) starting from the standard predicted PHDs for each target type.

\section{Probability generating functional (PGFL)} \label{Subsec:PGFL}

The probability generating functional is a convenient representation for stochastic modelling with a point process~\cite{Mah03}, a type of random process for which any one realisation consists of a set of isolated points either in time or space. Now, we model joint (probability generating) functionals which take into account the clutter due to the other target types in addition to the background for deriving the updated PHDs. Starting from the standard proved predicted PHDs in~\cite{Mah03} (refer to the appendix~\ref{sec:AppendixA} for the proof) for each multi-target type, we will derive novel extensions for the updated PHDs of N-type PHD filter from PGFLs for each target type for handling confusions between target types.

The joint functional for target type $i$ treating all other target types as clutter is given by

\begin{equation}
    F_i[g,h] = G_{T_i}(hG_{L_{i,i}}(g|.))G_{c_i}(g)\prod_{j=1\setminus i}^N G_{T_j}(G_{L_{j,i}}(g|.)),
\label{eqn:JointFunctional1}
\end{equation}
\noindent where $i \in \{1, ..., N\}$ denotes target type, $g$ is related to the target measurement process
and $h$ is related to the target state process.
%The joint functional for target type 2 treating target type 1 as clutter is given by
%\begin{equation}
%    F_2[g,h] = G_{T_2}(hG_{L_{2,2}}(g|.))G_{c_2}(g)G_{T_1}(G_{L_{1,2}}(g|.)),
%\label{eqn:JointFunctional2}
%\end{equation}
%\noindent
\begin{equation}
    G_{c_i}(h) = \exp(\lambda_i(c_i[g] - 1)),
\label{eqn:PGFLclutter1}
\end{equation}
\noindent where $G_{c_i}(h)$ is the Poisson PGFL~\cite{Mah03} for false alarms where $\lambda_i$ is the average number of false alarms for target type $i$ and the functional $c_i[g] = \int g(z)c_i(z)dz$ where $c_i(.)$ is the uniform density over the surveillance region;
\begin{equation}
    G_{T_i}(h) = \exp(\mu_i(s_i[g] - 1)),
\label{eqn:PGFLprior}
\end{equation}
\noindent where $G_{T_i}(h)$ is the prior PGFL and $\mu_i$ is the average number of targets, each of which is distributed according to $s_i(x)$ for target type $i$; and

\begin{equation}
    G_{L_{j,i}}(g|x) = 1 - p_{ji,D}(x) + p_{ji,D}(x)\int g(z) f_{ji}(z|x)dz,
\label{eqn:PGFLdetection}
\end{equation}
\noindent where $G_{L_{j,i}}(g|x)$ is the Bernoulli detection process for each target of target type $i$ using detector $j$ with probability of detection for target type $i$ by detector $j$, $p_{ji,D}$, and $f_{ji}(z|x)$ is a likelihood defining the probability that $z$ is generated by the target type $i$ conditioned on state $x$ from detector $j$~\cite{Mah03}. Expanding $s_i[hG_{L_{i,i}}(g|x)]$ and $s_j[G_{L_{j,i}}(g|x)]$ as
\begin{equation}
\begin{array} {lll}
    s_i[hG_{L_{i,i}}(g|x)] =& \displaystyle\int s_i(x) h(x)\Big(1 - p_{ii,D}(x) + \\&  p_{ii,D}(x)\int g(z) f_{ii}(z|x)dz\Big)dx,
\end{array}
\label{eqn:PGFLdetectionF1}
\end{equation}
\noindent and
\begin{equation}
\begin{array} {lll}
    s_j[G_{L_{j,i}}(g|x)] =&\displaystyle\int s_j(x)\Big(1 - p_{ji,D}(x) + \\&  p_{ji,D}(x)\int g(z) f_{ji}(z|x)dz\Big)dx,
\end{array}
\label{eqn:PGFLdetectionF2}
\end{equation}
\noindent Accordingly, $F_i[g,h]$ is expanded as
\begin{widetext}
\begin{equation}
\begin{array} {lll}
    F_i[g,h] =&  \exp \Bigg( \lambda_i \Big(\int g(z)c_i(z)dz - 1\Big) +  \sum_{j=1\setminus i}^N \mu_j \Big[\displaystyle\int s_j(x)\Big(1 - p_{ji,D}(x) + p_{ji,D}(x)\int g(z) f_{ji}(z|x)dz\Big)dx - 1 \Big] \\& + \mu_i \Big[\displaystyle\int s_i(x) h(x)\Big(1 - p_{ii,D}(x) + p_{ii,D}(x) \int g(z) f_{ii}(z|x)dz\Big)dx - 1 \Big] \Bigg) ,
\end{array}
\label{eqn:JointFunctional1Expanded}
\end{equation}
\end{widetext}
\noindent
%\noindent where $c[g] = \int g(z)c(z)dz$. Similarly, $F_2[g,h]$ is expanded as
%\begin{equation}
%\begin{array} {lll}
%    F_2[g,h] =& \exp(\lambda_2(c_{2}[g] - 1) + \mu_1[\int s_1(x)(1 - p_{12,D}(x) + \\& p_{12,D}(x)\int g(z) f_{12}(z|x)dz)dx - 1] + \mu_2[\int s_2(x) h(x)(1 - \\&p_{22,D}(x) + p_{22,D}(x) \int g(z) f_{22}(z|x)dz)dx - 1]) ,
%\end{array}
%\label{eqn:JointFunctional2Expanded}
%\end{equation}
%\noindent

The updated PGFL $G_i(h|z_1,...z_{M_j})$ for target type $i$ is obtained by finding the $M_j^{th}$ functional derivative of $F_i[g,h]$~\cite{Mah03} and is given by

\begin{equation}
    G_i(h|z_1,...,z_{M_j}) = \frac{\frac{\delta^{M_j}}{\delta_{\varphi_{z_1}}...\delta_{\varphi_{z_{M_j}}}} F_i[g,h]|_{g=0}} {\frac{\delta^{M_j}}{\delta_{\varphi_{z_1}}...\delta_{\varphi_{z_{M_j}}}} F_i[g,1]|_{g=0}},
\label{eqn:updatedPGFL}
\end{equation}
\noindent

The updated PHD for target type $i$ treating all other target types as clutter can be obtained by taking the first-order moment (mean)~\cite{Mah03} of Eq.~(\ref{eqn:updatedPGFL}) and setting $h = 1$,
\begin{widetext}
%\[
\begin{equation}
\begin{array} {lll}
\mathcal{D}_i(x|z_1,...,z_{M_j}) &= \frac{\delta}{\delta_{\varphi_x}} G_i(h|z_1,...z_{M_j})|_{h = 1}, \\
                             &= \mu_i s_i(x)(1 - p_{ii,D}(x)) +
\sum_{m=1}^{M_j} \frac{\mu_i s_i(x)p_{ii,D}(x)f_{ii}(z_m|x)}{\lambda_i c_i(z_m) + \sum_{j=1\setminus i}^N \mu_j \int s_j(x) p_{ji,D}(x) f_{ji}(z_m|x)dx + \mu_i\int s_i(x)p_{ii,D}(x)f_{ii}(z_m|x)dx},
\end{array}
\label{eqn:updatedPHD1}
\end{equation}
%\]
\end{widetext}
\noindent Thus, $\mathcal{D}_i(x|z_1,...,z_{M_j})$ in Eq.~(\ref{eqn:updatedPHD1}), is the updated PHD for target type $i$ treating all other target types as clutter. The term $\mu_i s_i(x)$ in Eq.~(\ref{eqn:updatedPHD1}) is the predicted PHD for target type $i$ (refer to the appendix~\ref{sec:AppendixA} for the proof). The following equational facts are important for the derivation of Eq.~(\ref{eqn:updatedPGFL}) to get Eq.~(\ref{eqn:updatedPHD1}) (refer to~\cite{Nat18}, Chapter 4, for the full derivation):

\begin{equation}
\begin{array} {lll}
 \frac{\delta}{\delta_{\varphi_x}} \lambda_i c_{i}(z_m) &= 0, \\ ~~ (\text{No~process~function~h(x)})\\
 \frac{\delta}{\delta_{\varphi_x}} \mu_j\int s_j(x) p_{ji,D}(x)f_{ji}(z_m|x)dx &= 0, \\ ~~ (\text{No~process~function~h(x))}\\
 \frac{\delta}{\delta_{\varphi_x}} \mu_i\int s_i(x) h(x) p_{ii,D}(x) f_{ii}(z_m|x)dx &= \\ ~~\mu_i s_i(x) p_{ii,D}(x) f_{ii}(z_m|x).
 \end{array}
\label{eqn:EquationalFacts}
\end{equation}
\noindent

\section{N-type PHD Filtering Strategy} \label{Subsec:DualPHDfilter}

Here we state PHD recursions in a generic form for multiple target, multiple type filtering with $Z_{1,k}, ... , Z_{N,k}$ separate but confused multi-target measurements between different target types, i.e. the N-type PHD filter, where $N \geq 2$. For N types of multiple targets, the PHDs, $\mathcal{D}_{\Xi_1}(x)$, $\mathcal{D}_{\Xi_2}(x)$, ..., $\mathcal{D}_{\Xi_N}(x)$, are the first-order moments of RFSs, $\Xi_1$, $\Xi_2$, ... $\Xi_N$, and they are intensity functions on a single state space $\mathcal{X}$ whose peaks identify the likely positions of the targets. For any region $R\subseteq \mathcal{X}$

\begin{equation}
    E[|(\Xi_1 \cup \Xi_2 ... \cup\Xi_N) \cap R|] = \sum_{i=1}^N  \int_R \mathcal{D}_{\Xi_i}(x)dx
\label{eqn:PHDcardinality}
\end{equation}

\noindent where$|.|$ is used to denote the cardinality of a set. In practice, Eq.(\ref{eqn:PHDcardinality}) means that by integrating the PHDs on any region $R$ of the state space, we obtain the expected number of targets (cardinality) in $R$.

%In this dual PHD filtering, if we denote $\mathcal{D}_{1,k|k}(x)$ and $\mathcal{D}_{2,k|k}(x)$  as the PHDs at time $k$ associated with the multi-target posterior density $p_{1,k|k}(X_{1,k}|Z_{1,1:k})$ and $p_{2,k|k}(X_{2,k}|Z_{2,1:k})$ for target type 1 and 2, respectively, then the Bayesian iterative prediction and update of $\mathcal{D}_{1,k|k}(x)$ and $\mathcal{D}_{2,k|k}(x)$ is known as the dual PHD filter which is a tractable alternative to the optimal intractable multi-target Bayes filter. %The recursion of the PHD filter is based on 3 assumptions:
%
%\begin{itemize}
%  \item The targets evolve and generate measurements independently of one another.
%  \item The clutter RFS, $K_k$, is poisson-distributed and is independent of target-originated measurements, and
%  \item The predicted multi-target RFS is poisson-distributed.
%\end{itemize}
%\noindent Though the first two assumptions are common to most Bayesian multi-target trackers~\cite{BarFor88}, the third is specific to the derivation of the PHD update operator where interactions amongst targets are negligible.
%
%Here, we model a dual PHD filter to filter two target types in such a way that one PHD update will filter the first target type treating the second as additional clutter (to background clutter), and the other PHD update will filter the second target type treating the first as additional clutter.

Accordingly, the Bayesian iterative prediction and update of the N-type PHD filtering strategy is given as follows.

The PHD $\textit{prediction}$ for target type $i$ is defined as
\begin{equation}
\begin{array} {lll}
    \mathcal{D}_{i,k|k-1}(x) = & \int p_{i,S,k|k-1}(\zeta)y_{i,k|k-1}(x|\zeta)\mathcal{D}_{i,k-1|k-1}(\zeta)d\zeta \\& + \gamma_{i,k}(x),
\end{array}
\label{eqn:PHDpredictioni}
\end{equation}
%\noindent
%The PHD $\textit{prediction}$ for target type 2 is defined as
%\begin{equation}
%    \mathcal{D}_{2,k|k-1}(x) = \int p_{s_2,k|k-1}(\zeta)y_{2,k|k-1}(x|\zeta)\mathcal{D}_{2,k-1|k-1}(\zeta)d\zeta + \gamma_{2,k}(x),
%\label{eqn:PHDprediction2}
%\end{equation}
\noindent where $\gamma_{i,k}(.)$ is the intensity function of a new target birth RFS $\Gamma_{i,k}$, $p_{i,S,k|k-1}(\zeta)$ is the probability that the target still exists at time $k$, $y_{i,k|k-1}(.|\zeta)$ is the single target state transition density at time $k$ given the previous state $\zeta$ for target type $i$.

Thus, following Eq.~(\ref{eqn:updatedPHD1}), the final updated PHD for target type $i$ is obtained by setting $\mu_is_i(x) = \mathcal{D}_{i,k|k-1}(x)$,

\begin{widetext}
\begin{equation}
\begin{array} {lll}  \mathcal{D}_{i,k|k}(x) &= \bigg[ 1 - p_{ii,D}(x) +
\sum_{z\in Z_{i,k}}\frac{p_{ii,D}(x)f_{ii,k}(z|x)}{c_{s_{i,k}}(z) + c_{t_{i,k}}(z) + \int p_{ii,D}(\zeta)f_{ii,k}(z|\zeta)\mathcal{D}_{i,k|k-1} (\zeta)d\zeta} \bigg] \mathcal{D}_{i,k|k-1(x)},
\end{array}
\label{eqn:PHDupdatei}
\end{equation}
\end{widetext}
\noindent The clutter intensity $c_{t_{i,k}}(z)$ due to all types of targets $j \in \{1, ..., N\}$ except target type $i$ in Eq.(\ref{eqn:PHDupdatei}) is given by

\begin{equation}
\begin{array} {lll}
    c_{t_{i,k}}(z) = \sum_{j \in \{1,...,N\}\setminus i} \int p_{ji,D}(y)\mathcal{D}_{j,k|k-1}(y)f_{ji,k}(z|y)dy,
\end{array}
\label{eqn:Clutteri}
\end{equation}
\noindent This means that when we are filtering target type $i$, all the other target types will be included as clutter. Eq.(\ref{eqn:Clutteri}) converts state space to observation space by integrating the PHD estimator $\mathcal{D}_{j,k|k-1}(y)$ and likelihood $f_{ji,k}(z|y)$ which defines the probability that $z$ is generated by detector $j$ conditioned on state $x$ of the target type $i$ taking into account the confusion probability $ p_{ji,D}(y)$, the detection probability for target type $i$ by detector $j$. Hence, it maps the state space of wrongly detected targets to the measurement space and treats them as clutter.

The clutter intensity due to the scene $i$, $c_{s_{i,k}}(z)$, in Eq.~(\ref{eqn:PHDupdatei}) is given by
\begin{equation}
    c_{s_{i,k}}(z) = \lambda_i c_i(z) = \lambda_{c_i} A c_i(z),
\label{eqn:Clutterscenei}
\end{equation}
\noindent where $c_i(.)$ is the uniform density over the surveillance region $A$, and $\lambda_{c_i}$ is the average number of clutter returns per unit volume for target type $i$ i.e. $\lambda_{i} = \lambda_{ci}A$. While the PHD filter has linear complexity with the current number of measurements ($m$) and with the current number of targets ($n$) i.e. computational order of $O(mn)$, the N-type PHD filter has linear complexity with the current number of measurements ($m$), with the current number of targets ($n$) and with the total number of target types ($N$) i.e. computational order of $O(mnN)$ as can be seen from Eq.~(\ref{eqn:PHDupdatei}).

In general, the clutter intensity due to the background for target type $i$, $c_{s_{i,k}}(z)$, can be different for each target type as they depend on the receiver operating characteristic (ROC) curves  of the detection processes. Moreover, the probabilities of detection $p_{ii,D}(x)$ and $p_{ji,D}(x)$ may all be different although assumed constant across both the time and space continua.

\begin{figure*}[!htb]%[!htb] %\begin{figure*}
\begin{center}
   \includegraphics[width=0.70\linewidth]{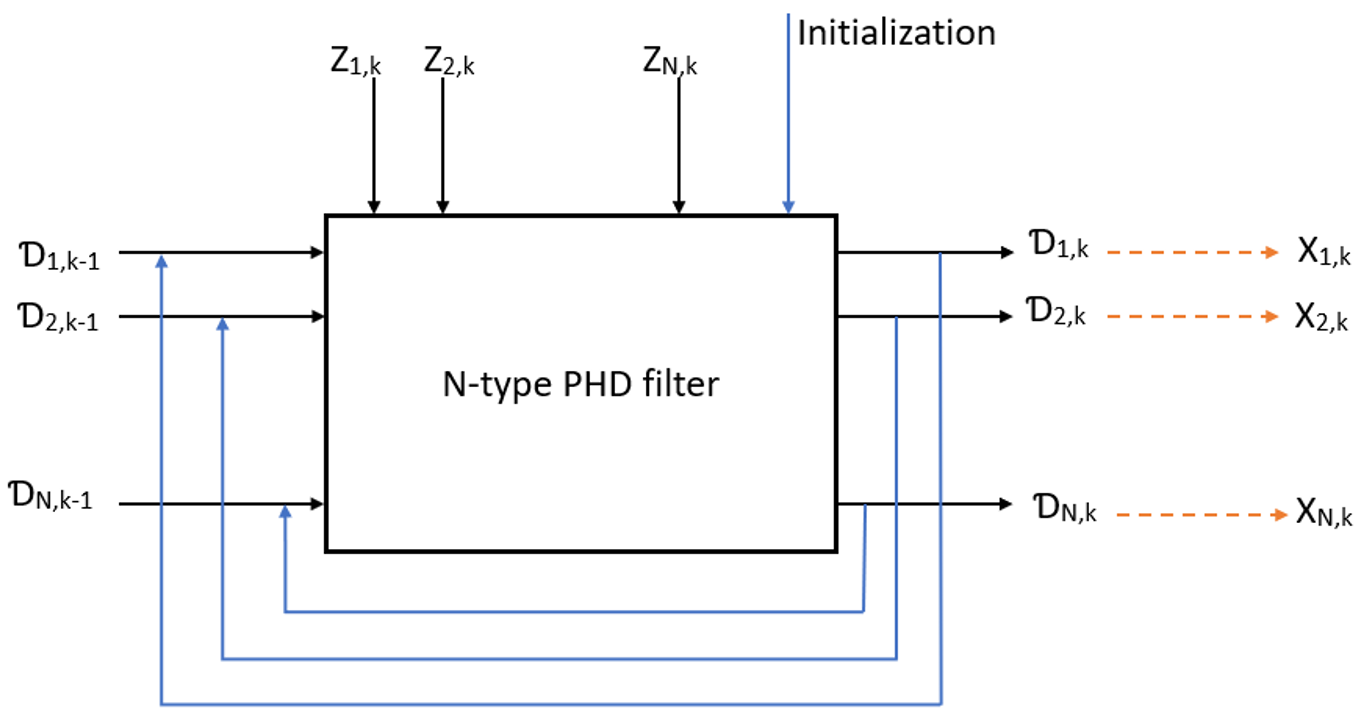}
\end{center}
   \caption{\small{Illustration of recursive N-type PHD filter structure.}} %\vspace{-5mm}
\label{fig:NtypePHDfilter}
\end{figure*}

The pictorial summary of recursive N-type PHD filter structure is given in Fig.~\ref{fig:NtypePHDfilter}. The intersection (common meeting point) of the N-type PHD filtering process is at Eq.~(\ref{eqn:Clutteri}) which involves the predicted PHDs and likelihoods of all other target types along with their corresponding probabilities of confusion when filtering target type $i$. This equation converts state space to observation space by integrating the PHD estimators (predicted PHDs) and likelihoods along with their corresponding probabilities of confusion to treat as clutter as clutter operates on the measurement space. Actually, Eq.~(\ref{eqn:Clutteri}) is called in Eq.~(\ref{eqn:PHDupdatei}) as a clutter along with a clutter due to the scene for updating PHD of target type $i$. At this point, the confusion of target types will be solved. Except this, every updated PHD retains its own type i.e. the updated PHDs are not combined. Initialization in Fig.~\ref{fig:NtypePHDfilter} shows the measurement-driven birth of targets. Thus, given the measurements of all target types at time $k$ and the updated PHDs at time $k-1$ ($\mathcal{D}_{i,k-1} = \mathcal{D}_{i,k-1|k-1}$ where $i \in \{1, ..., N\}$), the filtering process occurs recursively to get the updated PHDs at time $k$ ($\mathcal{D}_{i,k} = \mathcal{D}_{i,k|k}$ where $i \in \{1, ..., N\}$). Obviously, the updated PHDs at time $k-1$ are the PHDs after merging and pruning of the updated PHDs at the previous iteration (refer to section~\ref{Subsec:GM-DualPHD}) which will be predicted as existing targets in the next iteration (at time $k|k-1$). The estimated states are extracted from the updated PHDs of each target type (with weights greater than 0.5, refer to section~\ref{Subsec:GM-DualPHD}). In fact, after states extraction from the updated PHDs, the extracted states can be combined which do not affect the filtering process afterwards. However, for labelling of each target as in our preliminary work~\cite{BaiWal17}, keeping extracted states of each target type separately is important i.e. we apply the Hungarian algorithm to the extracted states of each target type separately for frame to frame labelling in~\cite{BaiWal17}, of course keeping the unique identities of all targets of the target types.

 % \end{figure*}
\noindent

\section{Gaussian Mixture-Based N-type PHD Filter Implementation} \label{Subsec:GM-DualPHD}

The Gaussian mixture implementation of the standard PHD (GM-PHD) filter~\cite{VoMa06} is a closed-form solution of the PHD filter with the assumptions of a linear Gaussian system.
In this section, this standard implementation is extended for the N-type PHD filter, more importantly solving Eq.~(\ref{eqn:Clutteri}).
We assume each target follows a linear Gaussian model.

\begin{equation}
    y_{i,k|k-1}(x|\zeta) =  \mathcal{N}(x;F_{i,k-1}\zeta, Q_{i,k-1})
\label{eqn:linearState1}
\end{equation}
\noindent
\begin{equation}
    f_{ji,k}(z|x) =  \mathcal{N}(z;H_{ji,k} x, R_{ji,k})
\label{eqn:linearObservation1}
\end{equation}
\noindent where $\mathcal{N}(.;m, P)$ denotes a Gaussian density with mean $m$ and covariance $P$; $F_{i,k-1}$ and $H_{ji,k}$ are the state transition and measurement matrices, respectively. $Q_{i,k-1}$ and $R_{ji,k}$ are the covariance matrices of the process and the measurement noises, respectively, where $i \in \{1,..., N\}$ and $j \in \{1,..., N\}$.
Besides, a current measurement driven birth intensity inspired by but not identical to~\cite{RisClaVoVo12} is introduced at each time step, removing the need for the prior knowledge (specification of birth intensities) or a random model, with a non-informative zero initial velocity. The intensity of the spontaneous birth RFS is $\gamma_{i,k}(x)$ for target type $i$

\begin{equation}
\begin{split}
     \gamma_{i,k}(x)  =  \sum_{v = 1}^{V_{\gamma_i,k}} w_{i,\gamma,k}^{(v)}\mathcal{N}(x; m_{i,\gamma,k}^{(v)}, P_{i,\gamma,k}^{(v)})
\label{eqn:PHDbirthassumption2}
\end{split}
\end{equation}
\noindent where $V_{\gamma_i,k}$ is the number of birth Gaussian components for target type $i$ where $i \in \{1, ..., N\}$, $m_{i,\gamma,k}^{(v)}$ is the current measurement (noisy version of position) and zero initial velocity used as mean and $P_{i,\gamma,k}^{(v)}$ is the birth covariance for a Gaussian component $v$ of target type $i$.

It is assumed that the posterior intensities for target type $i$ at time $k-1$ are Gaussian mixture of the form

\begin{equation}
\begin{split}
     \mathcal{D}_{i,k-1}(x)  =  \sum_{v = 1}^{V_{i,k-1}} w_{i,k-1}^{(v)}\mathcal{N}(x; m_{i,k-1}^{(v)}, P_{i,k-1}^{(v)}),
\label{eqn:PHDposterior1k-1}
\end{split}
\end{equation}

\noindent where $i \in \{1,..., N\}$ and $V_{i,k-1}$ is the number of Gaussian components of $\mathcal{D}_{i,k-1}(x)$. Under these assumptions, the predicted intensities at time $k$ for target type $i$ are given following Eq.~(\ref{eqn:PHDpredictioni}) by

\begin{equation}
    \mathcal{D}_{i,k|k-1}(x) = \mathcal{D}_{i,S,k|k}(x) + \gamma_{i,k}(x),
\label{eqn:PHDpredictionI1}
\end{equation}
\noindent where

\begin{equation}
\begin{array} {lll}  \mathcal{D}_{i,S,k|k-1}(x) =& p_{i,S,k} \sum_{v = 1}^{V_{i,k-1}} w_{i,k-1}^{(v)}\mathcal{N}(x; \\& m_{i,S,k|k-1}^{(v)},P_{i,S,k|k-1}^{(v)}), \nonumber
\end{array}
\label{eqn:PHDpredictionSurvival1}
\end{equation}
\noindent
\begin{equation}
 m_{i,S,k|k-1}^{(v)} = F_{i,k-1} m_{i,k-1}^{(v)},  \nonumber
\label{eqn:PHDpredictionSurvivalMean1}
\end{equation}
\noindent
\begin{equation}
 P_{i,S,k|k-1}^{(v)} = Q_{i,k-1} + F_{i,k-1} P_{i,k-1}^{(v)} F^T_{i,k-1},  \nonumber
\label{eqn:PHDpredictionSurvivalCov1}
\end{equation}
\noindent where $\gamma_{i,k}(x)$ is given by~(\ref{eqn:PHDbirthassumption2}).

Since $\mathcal{D}_{i,S,k|k-1}(x)$ and $\gamma_{i,k}(x)$ are Gaussian mixtures, $ \mathcal{D}_{i,k|k-1}(x)$ can be expressed as Gaussian mixture of the form

\begin{equation}
\begin{split}
     \mathcal{D}_{i,k|k-1}(x)  =  \sum_{v = 1}^{V_{i,k|k-1}} w_{i,k|k-1}^{(v)}\mathcal{N}(x; m_{i,k|k-1}^{(v)},P_{i,k|k-1}^{(v)}),
\label{eqn:PHDpredictionki}
\end{split}
\end{equation}
\noindent where $w_{i,k|k-1}^{(v)}$ is the weight accompanying the predicted Gaussian component $v$ for target type $i$ and $V_{i,k|k-1}$ is the number of predicted Gaussian components for target type $i$ where $i \in \{1,..., N\}$.

Now, assuming the probabilities of detection to be constant i.e. $p_{ii,D}(x) = p_{ii,D}$, the final updated PHD for target type $i$ is given as follows. Accordingly, the posterior intensity for target type $i$ at time $k$ (updated PHD) treating all other target types as clutter is also a Gaussian mixture which corresponds to Eq.~(\ref{eqn:PHDupdatei}), and is given by
\begin{equation}
\begin{split}
     \mathcal{D}_{i,k|k}(x)  =  (1 - p_{ii,D,k})\mathcal{D}_{i,k|k-1}(x) + \sum_{z\in Z_{i,k}} \mathcal{D}_{i,D,k}(x;z),
\label{eqn:PHDupdateki}
\end{split}
\end{equation}
\noindent where
\begin{equation}
\begin{split}
     \mathcal{D}_{i,D,k}(x;z)  =  \sum_{v = 1}^{V_{i,k|k-1}} w_{i,k}^{(v)}(z) \mathcal{N}(x; m_{i,k|k}^{(v)}(z), P_{i,k|k}^{(v)}), \nonumber
\label{eqn:PHDupdateDetki}
\end{split}
\end{equation}
\noindent
\begin{equation}
\begin{split}
     w^{(v)}_{i,k}(z)  =  \frac{p_{ii, D,k} w^{(v)}_{i,k|k-1} q^{(v)}_{i,k}(z)}{c_{s_{i,k}}(z) + c_{t_{i,k}}(z) + p_{ii, D,k} \sum_{l = 1}^{V_{i,k|k-1}} w^{(l)}_{i,k|k-1} q^{(l)}_{i,k}(z)}, \nonumber
\label{eqn:PHDupdatewwightki}
\end{split}
\end{equation}
\noindent
\begin{equation}
\begin{split}
     q^{(v)}_{i,k}(z)  =  \mathcal{N}(z; H_{ii,k} m_{i,k|k-1}^{(v)}, R_{ii,k} + H_{ii,k}P_{i,k|k-1}^{(v)} H^T_{ii,k}), \nonumber
\label{eqn:PHDupdateqki}
\end{split}
\end{equation}
\noindent
\begin{equation}
\begin{split}
     m^{(v)}_{i,k|k}(z)  =  m^{(v)}_{i,k|k-1} + K^{(v)}_{i,k} (z - H_{ii,k} m_{i,k|k-1}^{(v)}), \nonumber
\label{eqn:PHDupdatemki}
\end{split}
\end{equation}
\noindent
\begin{equation}
\begin{split}
     P^{(v)}_{i,k|k}  =  [I -  K^{(v)}_{i,k} H_{ii,k}] P_{i,k|k-1}^{(v)}, \nonumber
\label{eqn:PHDupdatepki}
\end{split}
\end{equation}
\noindent
\begin{equation}
\begin{split}
     K^{(v)}_{i,k}  =  P_{i,k|k-1}^{(v)} H^T_{ii,k} [ H_{ii,k}P_{i,k|k-1}^{(v)} H^T_{ii,k} + R_{ii,k}]^{-1}, \nonumber
\label{eqn:PHDupdateKki}
\end{split}
\end{equation}
\noindent $c_{s_{i,k}}(z)$ is given in Eq.~(\ref{eqn:Clutterscenei}). Therefore, all that left is to formulate the implementation scheme for $c_{t_{i,k}}(z)$ which is given in~(\ref{eqn:Clutteri}) and is given again as

\begin{equation}
\begin{array} {lll}
    c_{t_{i,k}}(z) = \sum_{j \in \{1,...,N\}\setminus i} \int p_{ji,D}(y)\mathcal{D}_{j,k|k-1}(y)f_{ji,k}(z|y)dy,
\end{array}
\label{eqn:Clutteri2}
\end{equation}
\noindent where $\mathcal{D}_{j,k|k-1}(y)$ is given in Eq.~(\ref{eqn:PHDpredictionki}), $f_{ji,k}(z|y)$ is given in Eq.~(\ref{eqn:linearObservation1}) and $p_{ji,D}(y)$ is assumed constant i.e. $p_{ji,D}(y) = p_{ji,D}$. Since $w_{j,k|k-1}^{(v)}$ is independent of the integrable variable $y$, Eq.~(\ref{eqn:Clutteri2}) becomes

\begin{equation}
\begin{array} {lll}
c_{t_{i,k}}(z) = & \sum_{j \in \{1,...,N\}\setminus i} \sum_{v = 1}^{V_{j,k|k-1}} p_{ji,D} w_{j,k|k-1}^{(v)} \int \mathcal{N}(y; \\& m_{j,k|k-1}^{(v)}, P_{j,k|k-1}^{(v)})\mathcal{N}(z;H_{ji,k} y, R_{ji,k})dy,
\label{eqn:Clutteri3}
\end{array}
\end{equation}
\noindent This can be simplified further using the following equality given that $P_1$ and $P_2$ are positive definite

\begin{equation}
\begin{array} {lll}
\int \mathcal{N} (y; m_1 \zeta, P_1)\mathcal{N}(\zeta; m_2, P_2)d\zeta =&  \mathcal{N}(y; m_1 m_2, \\& P_1 + m_1 P_2 m_2^T).
\end{array}
\label{eqn:Lemma1}
\end{equation}
\noindent Therefore, Eq.~(\ref{eqn:Clutteri3}) becomes,

\begin{equation}
\begin{array} {lll}
c_{t_{i,k}}(z) =& \sum_{j \in \{1,...,N\}\setminus i} \sum_{v = 1}^{V_{j,k|k-1}}  p_{ji,D}  w_{j,k|k-1}^{(v)} \mathcal{N}(z; \\& H_{ji,k} m_{j,k|k-1}^{(v)}, R_{ji,k} + H_{ji,k} P_{j,k|k-1}^{(v)}H_{ji,k}^T),
\label{eqn:Clutteri4}
\end{array}
\end{equation}
\noindent where $i \in \{1,..., N\}$.

\begin{algorithm}
\caption{Pseudocode for the N-type GM-PHD filter}\label{alg:tri-GMPHD}
\begin{algorithmic}[1]
\State \textbf{given} {\{$w_{i,k-1}^{(v)}, m_{i,k-1}^{(v)}, P_{i,k-1}^{(v)}\}_{v=1}^{V_{i,k-1}}$ for target type $i \in \{1, ..., N\}$, and the measurement set $Z_{j,k}$ for $j \in \{1, ..., N\}$}
\State \textbf{step 1.} {(prediction for birth targets)}
\For {$i = 1, ..., N$} \Comment{for all target type $i$}
\State $e_i = 0$
\For{$u = 1, ..., V_{\gamma_i,k}$}
\State $e_i := e_i + 1$
\State $w_{i,k|k-1}^{(e_i)} = w_{i,\gamma,k}^{(u)}$
\State $m_{i,k|k-1}^{(e_i)} = m_{i,\gamma,k}^{(u)}$
\State $P_{i,k|k-1}^{(e_i)} = P_{i,\gamma,k}^{(u)}$
\EndFor
\EndFor
\State \textbf{step 2.} {(prediction for existing targets)}
\For {$i = 1, ..., N$}  \Comment{for all target type $i$}
\For {$u = 1, ..., V_{i,k-1}$}
\State $e_i := e_i + 1$
\State $w_{i,k|k-1}^{(e_i)} = p_{i,S,k}w_{i,k-1}^{(u)}$
\State $m_{i,k|k-1}^{(e_i)} = F_{i,k-1}m_{i,k-1}^{(u)}$
\State $P_{i,k|k-1}^{(e_i)} = Q_{i,k-1} + F_{i,k-1} P_{i,k-1}^{(u)} F_{i,k-1}^T$
\EndFor
\EndFor
\State $V_{i,k|k-1} = e_i$
\State \textbf{step 3.} {(Construction of PHD update components)}
\For {$i = 1, ..., N$}  \Comment{for all target type $i$}
\For {$u = 1, ..., V_{i,k|k-1}$}
\State $\eta_{i,k|k-1}^{(u)} = H_{ii,k} m_{i,k|k-1}^{(u)}$
\State $S_{i,k}^{(u)} = R_{ii,k} + H_{ii,k}P_{i,k|k-1}^{(u)} H^T_{ii,k}$
\State $K^{(u)}_{i,k}  =  P_{i,k|k-1}^{(u)} H^T_{ii,k} [ S_{i,k}^{(u)} ]^{-1}$
\State $P^{(u)}_{i,k|k}  =  [I -  K^{(u)}_{i,k} H_{ii,k}] P_{i,k|k-1}^{(u)}$
\EndFor
\EndFor
\State \textbf{step 4.} {(Update)}
\For {$i = 1, ..., N$}  \Comment{for all target type $i$}
\For {$u = 1, ..., V_{i,k|k-1}$}
\State $w_{i,k}^{(u)} = (1 - p_{ii,D,k})w_{i,k|k-1}^{(u)}$
\State $m_{i,k}^{(u)} = m_{i,k|k-1}^{(u)}$
\State $P_{i,k}^{(u)} = P_{i,k|k-1}^{(u)}$
\EndFor
\State $l_i := 0$
\For{each $z \in Z_{j,k}$}
\State $l_i := l_i + 1$
\For {$u = 1, ..., V_{i,k|k-1}$}
\State $w_{i,k}^{(l_i V_{i,k|k-1} + u)} =~ p_{ii,D,k}w_{i,k|k-1}^{(u)}\mathcal{N}(z;
\myindent{4.2} \eta_{i,k|k-1}^{(u)}, S_{i,k}^{(u)}) $
%\Statex
\State $m_{i,k}^{(l_i V_{i,k|k-1} + u)} =~ m_{i,k|k-1}^{(u)} + K^{(u)}_{i,k} (z -
\myindent{4.3} \eta_{i,k|k-1}^{(u)})$
\State $P_{i,k}^{(l_i V_{i,k|k-1} + u)} = P^{(u)}_{i,k|k}$
\EndFor
\algstore{myalg}
\end{algorithmic}
\end{algorithm}

\begin{algorithm}
\begin{algorithmic} [1]                   % enter the algorithmic environment
\algrestore{myalg}
%\State \textbf{step 4.} {(Update)}
%\For {$i = 1, ..., N$}  \Comment{for all target type $i$}
%\For {$u = 1, ..., V_{i,k|k-1}$}
%\State $w_{i,k}^{(u)} = (1 - p_{ii,D,k})w_{i,k|k-1}^{(u)}$
%\State $m_{i,k}^{(u)} = m_{i,k|k-1}^{(u)}$
%\State $P_{i,k}^{(u)} = P_{i,k|k-1}^{(u)}$
%\EndFor
%\State $l_i := 0$
%\For{each $z \in Z_{j,k}$}
%\State $l_i := l_i + 1$
%\For {$u = 1, ..., V_{i,k|k-1}$}
%\State $w_{i,k}^{(l_i V_{i,k|k-1} + u)} =~ p_{ii,D,k}w_{i,k|k-1}^{(u)}\mathcal{N}(z;
%\myindent{4.2} \eta_{i,k|k-1}^{(u)}, S_{i,k}^{(u)}) $
%%\Statex
%\State $m_{i,k}^{(l_i V_{i,k|k-1} + u)} =~ m_{i,k|k-1}^{(u)} + K^{(u)}_{i,k} (z -
%\myindent{4.3} \eta_{i,k|k-1}^{(u)})$
%\State $P_{i,k}^{(l_i V_{i,k|k-1} + u)} = P^{(u)}_{i,k|k}$
%\EndFor
\For {$u = 1, ...., V_{i,k|k-1}$}
\State $c_{s_{i,k}}(z) = \lambda_{c_i} A c_i(z)$
\State $c_{t_{i,k}}(z) = \sum_{j \in \{1,...,N\}\setminus i} \sum_{e = 1}^{V_{j,k|k-1}}  p_{ji,D}  w_{j,k|k-1}^{(e)}
\myindent{1} \mathcal{N}(z; H_{ji,k} m_{j,k|k-1}^{(e)}, R_{ji,k} + H_{ji,k} P_{j,k|k-1}^{(e)}H_{ji,k}^T)$  \label{TriGMPHDconfusion}
\State $c_{i,k}(z) = c_{s_{i,k}}(z) + c_{t_{i,k}}(z)$
\State $w_{i,k,N} = \sum_{e=1}^{V_{i,k|k-1}} w_{i,k}^{(l_i V_{i,k|k-1} + e)}$
\State $w_{i,k}^{(l_i V_{i,k|k-1} + u)} = \frac{w_{i,k}^{(l_i V_{i,k|k-1} + u)}}{c_{i,k}(z) + w_{i,k,N}}$
\EndFor
\EndFor
\State $V_{i,k} = l_i V_{i,k|k-1} + V_{i,k|k-1}$
\EndFor
\State \textbf{output} {\{$w_{i,k}^{(v)}, m_{i,k}^{(v)}, P_{i,k}^{(v)}\}_{v=1}^{V_{i,k}}$}
%\BState
\end{algorithmic}
\label{alg:the_alg}
\end{algorithm}

The key steps of the N-type GM-PHD filter are summarised in Algorithms~\ref{alg:tri-GMPHD} and \ref{alg:tri-GMPHDprune}. The number of Gaussian components in the posterior intensities may increase without bound as time progresses. To keep the number of N-type GM-PHD components to a reasonable level after the measurement update, it is necessary to prune weak and duplicated components. First, weak components with weight $w_k^{v} < 10^{-5}$ are pruned. Further, Gaussian components with Mahalanobis distance less than $U = 4m$ from each other are merged. These pruned and merged Gaussian components, the output of Algorithm \ref{alg:tri-GMPHDprune}, will be predicted as existing targets in the next iteration. Finally, Gaussian components of the posterior intensity, the output of Algorithm \ref{alg:tri-GMPHD}, with means corresponding to weights greater than 0.5 as a threshold are selected as multi-target state estimates.

\begin{algorithm}
\caption{Pruning and merging for the N-type GM-PHD filter}\label{alg:tri-GMPHDprune}
\begin{algorithmic}[1]
\State \textbf{given} {\{$w_{i,k}^{(v)}, m_{i,k}^{(v)}, P_{i,k}^{(v)}\}_{v=1}^{V_{i,k}}$ for target type $i \in \{1, ..., N\}$, a pruning weight threshold T, and a merging distance threshold U.}
\For {$i = 1, ..., N$} \Comment{for all target type $i$}
\State  {Set $\ell_i = 0$, and $I_i = \{ v = 1, ..., V_{i,k}|w_{i,k}^{(v)} > T$ \}}
\State \textbf{repeat}
\State $\ell_i := \ell_i + 1$
\State $u := \arg\max_{v \in I_i} w_{i,k}^{(v)}$
\State $L_i := \Big\{v \in I_i \Big| (m_{i,k}^{(v)} - m_{i,k}^{(u)})^T (P_{i,k}^{(v)})^{-1} (m_{i,k}^{(v)} - m_{i,k}^{(u)}) \leq U \Big \}$
\State $\tilde{w}_{i,k}^{(\ell_i)} = \sum_{v \in L_i} w_{i,k}^{(v)}$
\State $\tilde{m}_{i,k}^{(\ell_i)} = \frac{1}{\tilde{w}_{i,k}^{(\ell_i)}} \sum_{v \in L_i} w_{i,k}^{(v)} m_{i,k}^{(v)}$
\State $\tilde{P}_{i,k}^{(\ell_i)} = \frac{1}{\tilde{w}_{i,k}^{(\ell_i)}} \sum_{v \in L_i} w_{i,k}^{(v)} (P_{i,k}^{(v)} + (\tilde{m}_{i,k}^{(\ell_i)} - m_{i,k}^{(v)})(\tilde{m}_{i,k}^{(\ell_i)} - m_{i,k}^{(v)})^T)$
\State $I_i := I_i \setminus L_i$
\State \textbf{until} {$I_i = \emptyset$}
\EndFor
\State \textbf{output} {\{$\tilde{w}_{i,k}^{(v)}, \tilde{m}_{i,k}^{(v)}, \tilde{P}_{i,k}^{(v)}\}_{v=1}^{\ell_i}$} as pruned and merged Gaussian components for target type $i$.

\end{algorithmic}
\end{algorithm}

\section{Experimental Results} \label{Sec:ExperimentalResults}

\begin{figure*}[!htb] %[t]%[!h]
  %\centering
  \begin{center}
%   \subfloat[\scriptsize{Output of 3 GM-PHD filters}]
  {\label{fig:SimIndep4Combined06} \includegraphics[width=1.0\textwidth]{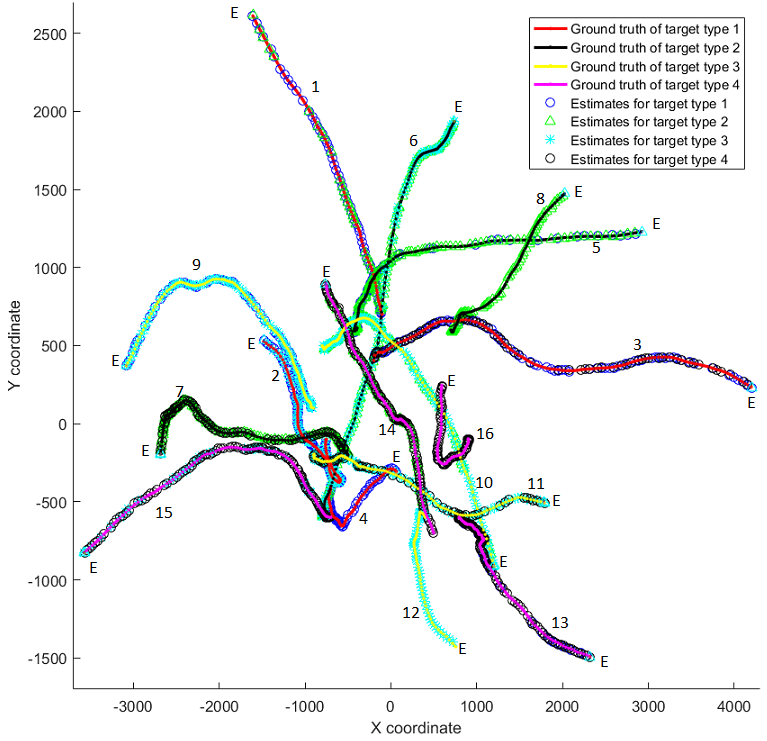}} \\ % 0.7,  N4IndepGM_PHD_Filter06.png, N4Indep_GM_PHD_06_new9.png
%   \subfloat[\scriptsize{Output of tri-GM-PHD filter}]
%  {\label{fig:SimTriCombined06} \includegraphics[width=0.70\textwidth]{triSim06.png}}
  \end{center}
   \caption{\small{Simulated ground truth (red, black, yellow and magenta for target type 1, 2, 3 and 4, respectively) and position estimates from four independent GM-PHD filters (blue circles, green triangles, cyan asterisks and black circles for target type 1, 2, 3 and 4, respectively) using $p_{12,D} = p_{13,D}= p_{14,D} = p_{21,D} = p_{23,D} = p_{24,D} = p_{31,D} = p_{32,D} = p_{34,D} = p_{41,D}  = p_{42,D} = p_{43,D} = 0.6$.}} %\vspace{-5mm}
  \label{fig:IndependentComparison42}
\end{figure*}

\begin{figure*}[!htb] %[t]%[!h]
  %\centering
  \begin{center}
%   \subfloat[\scriptsize{Output of 3 GM-PHD filters}]
%  {\label{fig:SimIndep3Combined06} \includegraphics[width=0.70\textwidth]{ThreephdSim06.png}} \\ % 0.7
%   \subfloat[\scriptsize{Output of tri-GM-PHD filter}]
  {\label{fig:SimQuadCombined06} \includegraphics[width=1.0\textwidth]{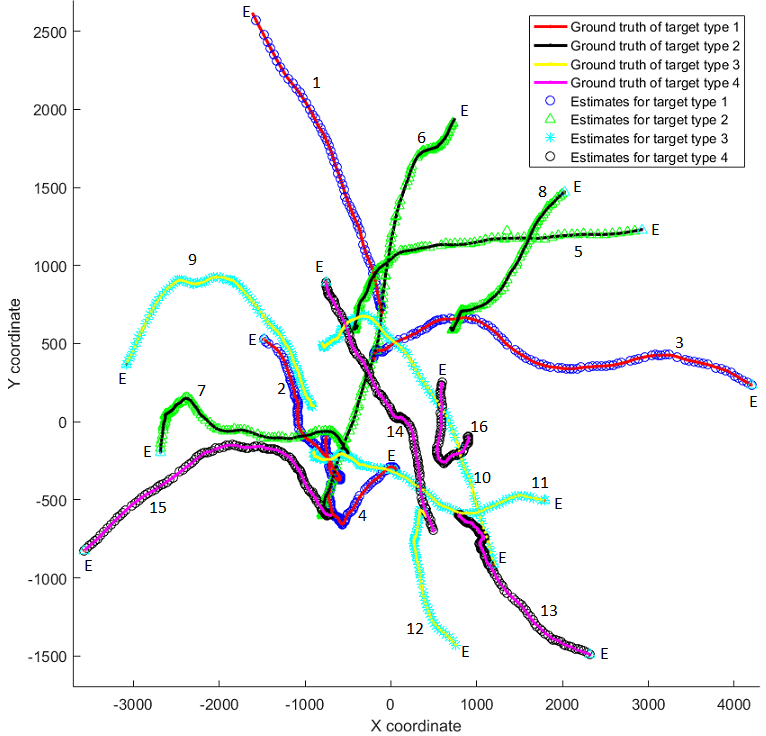}}%Quad_GM_PHD_Filter06.png,Quad_GM_PHD_06_new17.png
  \end{center}
   \caption{\small{Simulated ground truth (red, black, yellow and magenta for target type 1, 2, 3 and 4, respectively) and position estimates from quad GM-PHD filter (blue circles, green triangles, cyan asterisks and black circles for target type 1, 2, 3 and 4, respectively) using $p_{12,D} = p_{13,D}= p_{14,D} = p_{21,D} = p_{23,D} = p_{24,D} = p_{31,D} = p_{32,D} = p_{34,D} = p_{41,D}  = p_{42,D} = p_{43,D} = 0.6$.}} %\vspace{-5mm}
  \label{fig:QuadComparison42}
\end{figure*}

In this section, a simulation filtering example using a quad GM-PHD filter for four different types of multiple targets is analyzed. We have also made experimental simulation analyses of a dual GM-PHD filter ($N=2$) and a tri-GM-PHD filter ($N=3$) and then applied the tri-GM-PHD filter for visual tracking applications in~\cite{BaiWal17} for three types of targets. In this experiment, we demonstrate the quad GM-PHD filter (N = 4) with detailed analysis as a typical simulation example under different values of confusion detection probabilities. Accordingly, we define a sequence of 120 frames with sixteen trajectories that emanate from four types of targets that appear in the scene at different positions and time steps (frames), as shown in Fig.~\ref{fig:IndependentComparison42} and Fig.~\ref{fig:QuadComparison42}. This is a typical example of not only a higher number of target types (four) but also an example of a dense scene i.e. it consists of trajectories of 16 targets in the same scene with many crossings. Our results are obtained after running 50 simulations i.e. the number of Monte Carlo (MC) is 50. Obviously, the goal of a N-type PHD filter is to handle confusions among $N \geq 2$ different target types; not to deal with sparse or dense targets in the scene. With regards to sparse or dense targets in the scene, it has the same characteristics as the standard PHD filter.

%The initial locations and covariances for all target types are given by Eq.~\ref{eqn:PHDbirthSimulation41} and Eq.~\ref{eqn:PHDbirthSimulation42} as follows.

The initial locations and covariances for all target types are given by Eq.~(\ref{eqn:PHDbirthSimulation41}) as follows.
\clearpage
\noindent
\begin{equation}
\begin{split}
           m_{1,k}^{(1)} &= [-100, 700, 0, 0]^T, \\
           m_{1,k}^{(2)} &= [-750, -100, 0, 0]^T, \\
           m_{1,k}^{(3)} &= [-200, 400, 0, 0]^T, \\
           m_{1,k}^{(4)} &= [-700, -400, 0, 0]^T,\\
           m_{2,k}^{(5)} &= [-400, 600, 0, 0]^T, \\
           m_{2,k}^{(6)} &= [-800, -600, 0, 0]^T, \\
           m_{2,k}^{(7)} &= [-500, -200, 0, 0]^T, \\
           m_{2,k}^{(8)} &= [700, 600, 0, 0]^T, \\
           m_{3,k}^{(9)} &= [-900, 100, 0, 0]^T,  \\
           m_{3,k}^{(10)} &= [-800, 500, 0, 0]^T,  \\
           m_{3,k}^{(11)} &= [-900, -200, 0, 0]^T,  \\
           m_{3,k}^{(12)} &= [400, -600, 0, 0]^T,  \\
           m_{4,k}^{(13)} &= [800, -600, 0, 0]^T,  \\
           m_{4,k}^{(14)} &= [500, -700, 0, 0]^T,  \\
           m_{4,k}^{(15)} &= [-700, -600, 0, 0]^T,  \\
           m_{4,k}^{(16)} &= [900, -100, 0, 0]^T,  \\
           P_{1,k} = P_{2,k} = P_{3,k} &= P_{4,k} = diag([200, 200, 100, 100]).
%          P_{1,k} &= P_{2,k} = diag([100, 100, 25, 25]).
\label{eqn:PHDbirthSimulation41}
\end{split}
\end{equation}
\noindent

%\begin{equation}
%\begin{split}
%%           m_{1,k}^{(1)} &= [-100, 700, 0, 0]^T, \\
%%           m_{1,k}^{(2)} &= [-750, -100, 0, 0]^T, \\
%%           m_{1,k}^{(3)} &= [-200, 400, 0, 0]^T, \\
%%           m_{1,k}^{(4)} &= [-700, -400, 0, 0]^T,\\
%%           m_{2,k}^{(5)} &= [-400, 600, 0, 0]^T, \\
%%           m_{2,k}^{(6)} &= [-800, -600, 0, 0]^T, \\
%%           m_{2,k}^{(7)} &= [-500, -200, 0, 0]^T, \\
%%           m_{2,k}^{(8)} &= [700, 600, 0, 0]^T, \\
%           m_{3,k}^{(9)} &= [-900, 100, 0, 0]^T,  \\
%           m_{3,k}^{(10)} &= [-800, 500, 0, 0]^T,  \\
%           m_{3,k}^{(11)} &= [-900, -200, 0, 0]^T,  \\
%           m_{3,k}^{(12)} &= [400, -600, 0, 0]^T,  \\
%           m_{4,k}^{(13)} &= [800, -600, 0, 0]^T,  \\
%           m_{4,k}^{(14)} &= [500, -700, 0, 0]^T,  \\
%           m_{4,k}^{(15)} &= [-700, -600, 0, 0]^T,  \\
%           m_{4,k}^{(16)} &= [900, -100, 0, 0]^T,  \\
%           P_{3,k} &= P_{4,k} = diag([100, 100, 25, 25]).
%\label{eqn:PHDbirthSimulation42}
%\end{split}
%\end{equation}
%\noindent

The state vector $x_k= [p_{x,xk}, p_{y,xk},\dot{p}_{x,xk},\dot{p}_{y,xk}]^T$  consists of position $(p_{x,xk}, p_{y,xk})$ and velocity $(\dot{p}_{x,xk}, \dot{p}_{y,xk})$, and the measurement is a noisy version of the position, $z_k = [p_{x,zk}, p_{y,zk}]^T$.
Each of the target trajectories follows a linear Gaussian dynamic model of Eq.~(\ref{eqn:linearState1}) with matrices

\[ F_{i,k-1} = \left[ \begin{array}{cc}
           I_2 & \Delta I_2 \\
           0_2 & I_2
           \end{array} \right], \]
%\[ Q_{i, k-1} = \sigma_{v_i}^2 \left [ \begin{array}{cc}
%    \frac{\Delta^4}{4}I_2 & \frac{\Delta^3}{2}I_2 \\
%    \frac{\Delta^3}{2}I_2 & \Delta^2 I_2
%    \end{array} \right] \]
\begin{equation} Q_{i, k-1} = \sigma_{v_i}^2 \left [ \begin{array}{cc}
    \frac{\Delta^4}{4}I_2 & \frac{\Delta^3}{2}I_2 \\
    \frac{\Delta^3}{2}I_2 & \Delta^2 I_2
    \end{array} \right],
\label{eqn:PHDstateTransitionMatrix}
\end{equation}
\noindent where $I_n$ and $0_n$ denote the $n \times n$ identity and zero matrices, respectively. $\Delta = 1s$ is the sampling period. $\sigma_{v_i} = 5m/s^2$  where $i \in \{1,2,3,4\}$ is the standard deviation of the process noise for target type $i$.
%$F_{1,k-1}=[I_2, \Delta I_2; 0_2, I_2]$, $F_{2,k-1} =[I_2, \Delta I_2; 0_2, I_2]$, $Q_{1, k-1}=\sigma_{v1}^2[\Delta^4I_2/4, \Delta^3I_2/2; \Delta^3I_2/2, \Delta^2 I_2]$, $Q_{2, k-1}=\sigma_{v2}^2[\Delta^4I_2/4, \Delta^3I_2/2; \Delta^3I_2/2, \Delta^2 I_2]$ where $I_n$ and $0_n$ denotes the \textit{n} x \textit{n} identity and zero matrices, respectively. $\Delta = 1s$ is the sampling period. $\sigma_{v1}^2 = \sigma_{v2}^2 = 5m/s^2$ are the standard deviations of the process noise.

For the algorithm, we assume each target has a survival probability $p_{1,S} = p_{2,S} = p_{3,S} = p_{4,S} = 0.99$. The probabilities of detection are $p_{11,D} = 0.90$ , $p_{22,D} = p_{33,D} = 0.92$, $p_{44,D} = 0.91$, and different values of confusion detection probabilities (0.0, 0.3, 0.6 and 0.9) are analyzed for $p_{12,D}$, $p_{13,D}$, $p_{14,D}$, $p_{21,D}$, $p_{23,D}$, $p_{24,D}$, $p_{31,D}$, $p_{32,D}$, $p_{34,D}$, $p_{41,D}$, $p_{42,D}$ and $p_{43,D}$.

The measurement follows the observation model of Eq.~(\ref{eqn:linearObservation1}) with matrices

\begin{equation}
\begin{split}
     H_{ii,k} = H_{ji,k} =  [I_2~~0_2], \\
    R_{ii,k} = \sigma_{r_{ii}}^2 I_2,  \\
    R_{ji,k} = \sigma_{r_{ji}}^2 I_2,
\label{eqn:PHDobservationMatrix}
\end{split}
\end{equation}
\noindent where $\sigma_{r_{ii}} = \sigma_{r_{ji}} = 6m$ ($i \in \{1,2,3,4\}$ and $j \in \{1,2,3,4\}$) is the standard deviation of the measurement noise.
%The measurement follows the observation models with matrices of $H_{11,k}=H_{22,k}=H_{12,k}=H_{21,k}=[I_2, 0_2]$, $R_{11,k}=\sigma_{r11}^2 I_2$, $R_{22,k}=\sigma_{r22}^2 I_2$, $R_{12,k}=\sigma_{r12}^2 I_2$, and $R_{21,k}=\sigma_{r21}^2 I_2$ where $\sigma_{r11} = \sigma_{r22} = \sigma_{r12}=\sigma_{r21}=6m$ is the standard deviation of the measurement noise.

%Similar to the ones in section~\ref{Sec:DualExperimentalSimResults} and section~\ref{Sec:TriExperimentalSimResults}, the state vector $x_k= [p_{x,xk}, p_{y,xk},\dot{p}_{x,xk},\dot{p}_{y,xk}]^T$  consists of position $(p_{x,xk}, p_{y,xk})$ and velocity $(\dot{p}_{x,xk}, \dot{p}_{y,xk})$, and the measurement is a noisy version of the position, $z_k = [p_{x,zk}, p_{y,zk}]^T$. Each of the target trajectories follows a linear Gaussian dynamic model of Eq.~(\ref{eqn:linearState1}) with matrices given in Eq.~(\ref{eqn:PHDstateTransitionMatrix}). In this case we set $\Delta = 1s$ as the sampling period and $\sigma_{v_i} = 5m/s^2$  where $i \in \{1,2,3,4\}$ as the standard deviation of the process noise for target type $i$.

Since there are many targets in the scene, we use about 40 clutter returns (10 for each target type) over the surveillance region. A current measurement driven birth intensity inspired by but not identical to~\cite{RisClaVoVo12} is introduced at each time step, removing the need for the prior knowledge (specification of birth intensities) or a random model, with a non-informative zero initial velocity. At birth, Gaussian components of each target type have a corresponding initial weight $ w_{1,\gamma,k}^{(i)} = w_{2,\gamma,k}^{(i)} = w_{3,\gamma,k}^{(i)} = w_{4,\gamma,k}^{(i)}= 3\times10^{-6}$. This very small initial weight is assigned to the Gaussian components for new births as this is effective for high clutter rates. This is basically equivalent to the average number of appearing (birth) targets per time step ($n_b$) divided uniformly across the surveillance region ($A$).

\begin{figure*}[!htb]%[!htb] %\begin{figure*}
\begin{center}
   \includegraphics[width=1.0\linewidth]{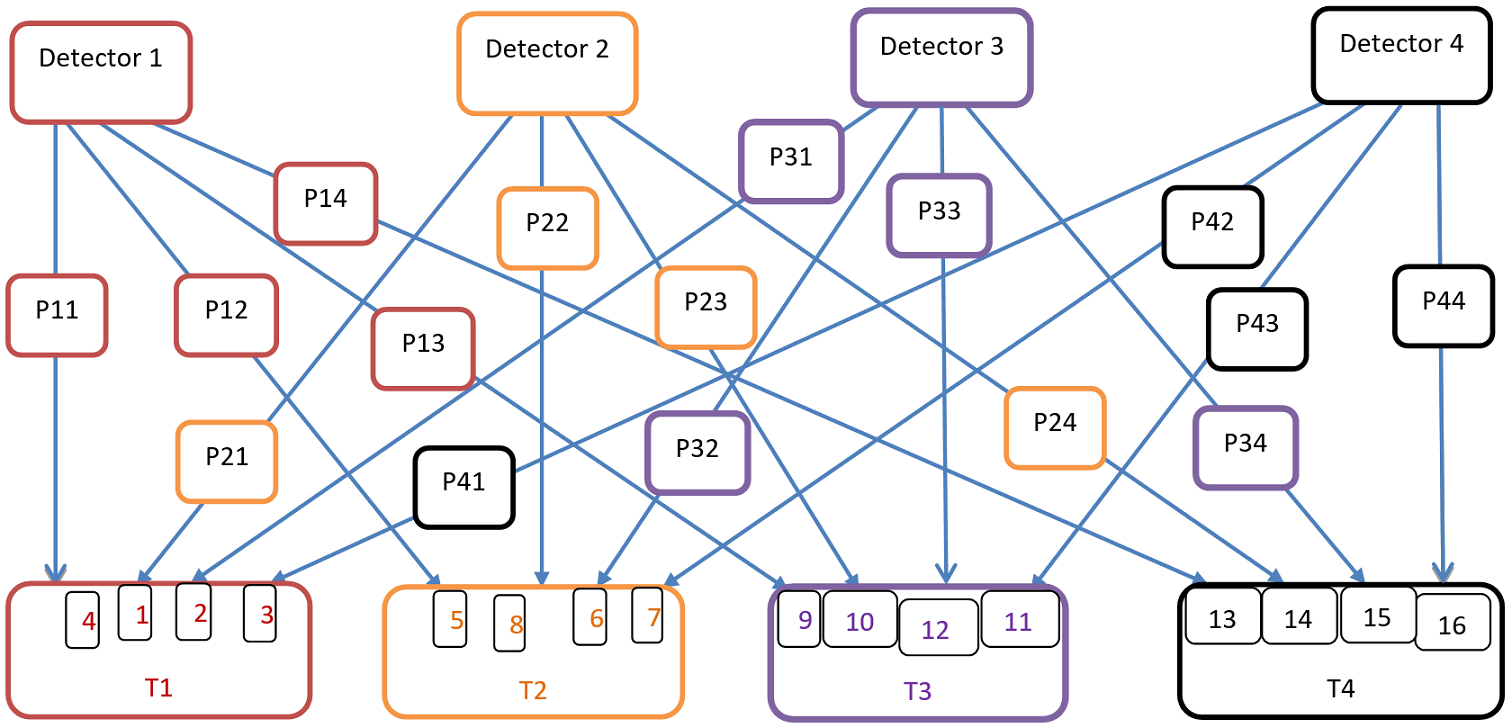}
\end{center}
   \caption{\small{Confusions between four target types (T1, T2, T3 and T4) at the detection stage from detectors 1, 2, 3 and 4.}} %\vspace{-5mm}
\label{fig:ConfusionsPs4}
\end{figure*}

The configuration of the detectors is shown Fig.~\ref{fig:ConfusionsPs4}. As shown in this figure, detector 1 detects target type 1 (targets 1, 2, 3 and 4) with probability of detection $p_{11,D}$, target 5 which is of target type 2 with probability of detection $p_{12,D}$, target 9 which is of target type 3 with probability of detection $p_{13,D}$ and target 13 which is of target type 4 with probability of detection $p_{14,D}$. Detector 2 detects target type 2 (targets 5, 6, 7 and 8) with probability of detection $p_{22,D}$, target 1 which is of target type 1 with probability of detection $p_{21,D}$, target 10 which is of target type 3 with probability of detection $p_{23,D}$ and target 14 which is of target type 4 with probability of detection $p_{24,D}$. Similarly, detector 3 detects target type 3 (targets 9, 10, 11 and 12) with probability of detection $p_{33,D}$, target 2 which is of target type 1 with probability of detection $p_{31,D}$, target 6 which is of target type 2 with probability of detection $p_{32,D}$ and target 15 which is of target type 4 with probability of detection $p_{34,D}$. Moreover, detector 4 detects target type 4 (targets 13, 14, 15 and 16) with probability of detection $p_{44,D}$, target 3 which is of target type 1 with probability of detection $p_{41,D}$, target 7 which is of target type 2 with probability of detection $p_{42,D}$ and target 11 which is of target type 3 with probability of detection $p_{43,D}$. This means that targets 1, 2 and 3 from target type 1, targets 5, 6 and 7 from target type 2, targets 8, 9 and 10 from target type 3, and targets 13, 14 and 15 from target type 4 are detected two times. Our main goal is to filter out confused measurements which correspond to a specific target i.e. doubly detected targets are estimated once, not twice. Therefore, the number of targets in the scene is 16 maximum, not 28.

 % \end{figure*}
\noindent

The figures shown in Fig.~\ref{fig:IndependentComparison42} and Fig.~\ref{fig:QuadComparison42} show the comparisons of the outputs of both the quad GM-PHD filter (Fig.~\ref{fig:QuadComparison42}) and four independent GM-PHD filters (Fig.~\ref{fig:IndependentComparison42}) for detection probability of confusion of 0.6. For both approaches, the simulated ground truths are shown in red for target type 1, black for target type 2, yellow for target type 3 and magenta for target type 4 while the estimates are shown in blue circles for target type 1, green triangles for target type 2, cyan asterisks for target type 3 and black circles for target type 4. Accordingly, for simulated measurements, the quad GM-PHD filter outputs estimates of target type 1 (targets 1, 2, 3 and 4), target type 2 (targets 5, 6, 7 and 8), target type 3 (targets 9, 10, 11 and 12) and target type 4 (targets 13, 14, 15 and 16) being well differentiated %which can not be handled even by intelligent track management
as shown in Fig.~\ref{fig:QuadComparison42}. However, using four independent GM-PHD filters, estimates of targets 1, 2, 3, 4, 5, 9 and 13 are obtained from GM-PHD filter 1, estimates of targets 1, 5, 6, 7, 8, 10 and 14 from GM-PHD filter 2, estimates of targets 2, 9, 10, 11, 12 and 15 from GM-PHD filter 3, and estimates of targets 3, 7, 11, 13, 14, 15 and 16 from GM-PHD filter 4 i.e. targets 1, 2, 3, 5, 6, 7, 9, 10, 11, 13, 14 and 15 are estimated twice though intermittently, as shown in Fig.~\ref{fig:IndependentComparison42} overlayed. Even if the confusion rates increase to 0.6, the proposed method filters out the target confusions effectively, discriminating the target types. This confusion problem is solved by using our proposed approach with a computation time of 251.27 seconds for 120 iterations when compared to 152.06 seconds using four independent GM-PHD filters implemented on a Core i7 2.30 GHz processor and a 8 GB RAM laptop using MATLAB when setting detection probabilities of confusion to 0.6, for example, as given in Table~\ref{tbl:OSPAerrorSim4}. In Fig.~\ref{fig:IndependentComparison42} and Fig.~\ref{fig:QuadComparison42}, E stands for the end point whereas the other side of each target is the starting point of the simulation.

%Similar to the dual PHD and tri-PHD filters case, the quad-GM-PHD filter degrades to four GM-PHD filters when we set the probabilities of confusion to 0.0 i.e. no target confusions.
When we set the probabilities of confusion to 0.0 i.e. no target confusions, the quad GM-PHD filter performs similar to four GM-PHD filters i.e. it degrades to four GM-PHD filters.  However, if the values of confusion detection probabilities are very close to the values of the true detection probabilities, the quad GM-PHD filter still effectively filters the confusion in the detection of targets though it sometimes fails to discriminate the target types. If the probability of confusion is the same as of true detection, then the result is random on first guess (sometimes fails to discriminate the target types) though it still filters out the confusions effectively. Therefore, the values of the confusion probabilities ($p_{12,D}$, $p_{13,D}$, $p_{14,D}$, $p_{21,D}$, $p_{23,D}$, $p_{24,D}$, $p_{31,D}$, $p_{32,D}$, $p_{34,D}$, $p_{41,D}$, $p_{42,D}$ and $p_{43,D}$) should be less than the values of the true detection probabilities ($p_{11,D}$, $p_{22,D}$, $p_{33,D}$ and $p_{44,D}$) to discriminate the target types. However, in real applications (e.g. visual tracking), this does not happen, i.e. the confusion detection probabilities can never become equal in values to the true detection probabilities as object detectors are at least becoming more accurate than random guessing; nobody would employ a random detector.

On the other hand, if each target is regarded as a type (e.g. each of the four target types in this example has only one target), the N-type GM-PHD filter is used as a labeler of each target i.e. it discriminates those targets from frame to frame (from one time step to another) whether or not confusions between targets exist rather than simply degrading to the standard GM-PHD filter(s). Obviously, the number of target types ($N \geq 2$) needs to be known in advance so that measurements from each target type will be collected and then provided to the algorithm.

\noindent
\noindent

\begin{figure*}[!htb] %[!htb] %[t]%[!h]
  %\centering
  \begin{center}
   \subfloat[\scriptsize{Cardinality}]
  {\label{fig:SimCardinality4} \includegraphics[width=0.5\textwidth]{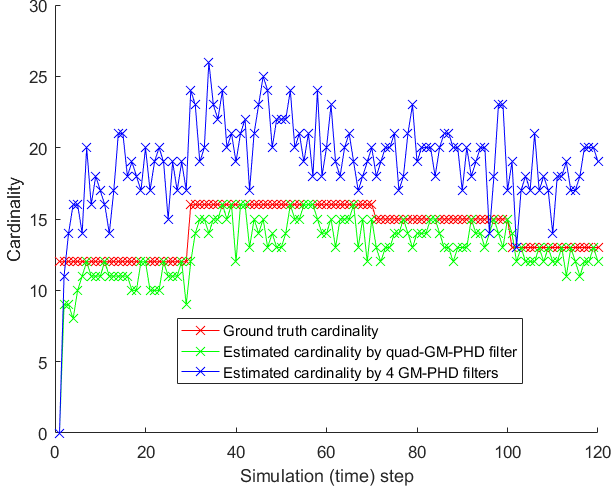}} %\\ %width=0.5 SimCardinalityn4.png
   \subfloat[\scriptsize{OSPA error}]
  {\label{fig:SimOSPAerror4} \includegraphics[width=0.5\textwidth]{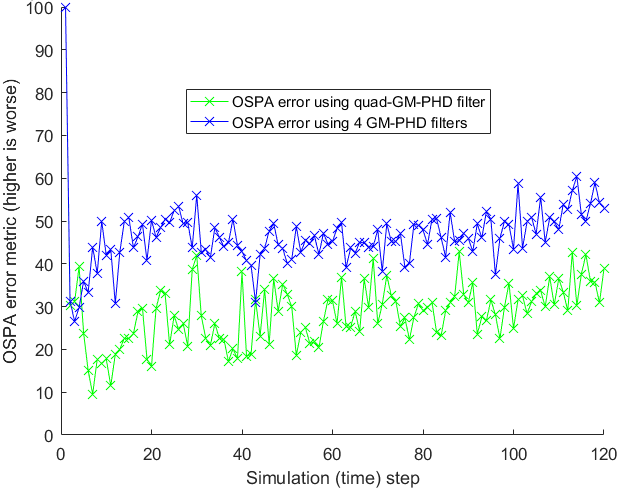}} % SimOSPAn4.png
  \end{center}
   \caption{\small{Cardinality and OSPA error: Ground truth (red for cardinality only), quad GM-PHD filter (green), four independent GM-PHD filters (blue) for $p_{12,D} = p_{13,D}= p_{14,D} = p_{21,D} = p_{23,D} = p_{24,D} = p_{31,D} = p_{32,D} = p_{34,D} = p_{41,D}  = p_{42,D} = p_{43,D} = 0.6$.}} \vspace{-5mm}
  \label{fig:OSPA-CardinalityComparison4}
\end{figure*}

\begin{figure*}[!htb]%[!htb] %\begin{figure*}
\begin{center}
    \includegraphics[width=.50\linewidth]{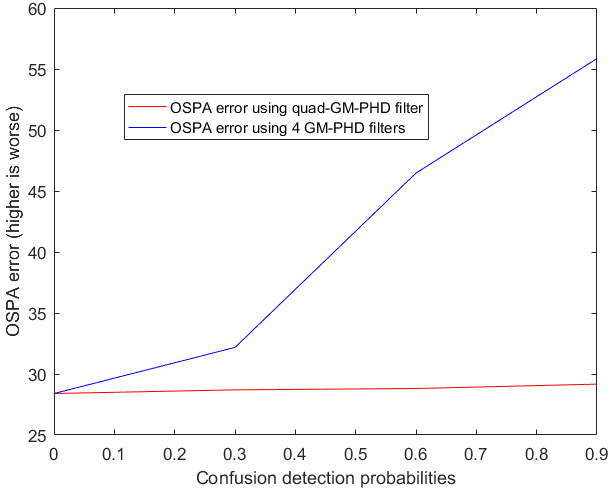}
\end{center}
   \caption{\small{OSPA error comparison for quad GM-PHD filter and four independent GM-PHD filters at different probabilities of confusion.}} %\vspace{-5mm}
\label{fig:OSPAcomparisonN4}
\end{figure*}

Furthermore, we assess tracking accuracy using the cardinality (number of targets) and Optimal Subpattern Assignment (OSPA) metric~\cite{SchVoVo08} (using order $p = 1$ and cutoff $c = 100$). From Fig.~\ref{fig:SimCardinality4} (when setting the probabilities of confusion to 0.6), we observe that the cardinality of targets estimated using four independent GM-PHD filters (blue) has much more deviation from the ground truth (in red) when compared to the one obtained using our proposed quad GM-PHD filter (green). Similarly, the OSPA error of using four independent GM-PHD filters (blue) is much greater than that of using the quad GM-PHD filter (green) as shown in Fig.~\ref{fig:SimOSPAerror4}. The overall average value of the OSPA error for four independent GM-PHD filters is 46.47m compared to 28.81m when using our proposed quad GM-PHD filter as given in Table~\ref{tbl:OSPAerrorSim4}. The OSPA error and time taken (in brackets) when using probabilities of confusion of 0.3 and 0.9 are also given in Table~\ref{tbl:OSPAerrorSim4}. As can be observed from Fig.~\ref{fig:OSPAcomparisonN4} and Table~\ref{tbl:OSPAerrorSim4}, as we increase the probabilities of confusions from 0.0 to 0.9, the OSPA error for quad GM-PHD filter is almost constant which shows how efficient the quad GM-PHD filter is in handling target confusions. However, for the case of using four independent GM-PHD filters, the OSPA error increases significantly as we increase the probabilities of confusions from 0.0 to 0.9 which is due to the increase in target confusions. The time taken (given in Table~\ref{tbl:OSPAerrorSim4}) also increases slightly for both methods with the increment of target confusions.

\noindent
%\clearpage

\begin{table*}[!htb]%[!h]%[tb]
\begin{center}
%\resizebox{1.0\textwidth}{!}{\begin{minipage}{\textwidth}
\begin{tabular}{|l|c|c|r|}
\hline
Method & 0.3 & 0.6 & 0.9 \\
\hline\hline
Quad GM-PHD filter & 28.70m (239.82sec) & 28.81m (251.27sec)& 29.17m (270.13sec)\\
4 GM-PHD filters & 32.18m (147.42sec) & 46.47m (152.06sec)& 55.86m (160.17sec)\\
\hline
\end{tabular}
%\end{minipage}}
\end{center}
\caption{\small{OSPA error at different values of probabilities of confusion $p_{12,D}$, $p_{13,D}$, $p_{14,D}$, $p_{21,D}$, $p_{23,D}$, $p_{24,D}$, $p_{31,D}$, $p_{32,D}$, $p_{34,D}$, $p_{41,D}$, $p_{42,D}$ and $p_{43,D}$ (0.3, 0.6 and 0.9) for quad GM-PHD filter and 4 independent GM-PHD filters. Time taken is given in brackets.}}
\label{tbl:OSPAerrorSim4}
\end{table*}
\noindent

 % \end{figure*}
\noindent

 % corresponding to color mypink

\section{Conclusion} \label{Sec:Conclusion}

In this work, we propose a novel, N-type PHD filter where $N\geq2$, which is an extension of the standard PHD filter in the RFS framework to account for many different types of targets with separate but confused observations of the same scene. In this approach, we assume that there are confusions between detections, i.e. clutter arises not just from background false positives, but also from target confusion. Under the Gaussianity and linearity assumptions, the Gaussian mixture (GM) implementation is proposed for this N-type PHD filter, N-type GM-PHD filter. We evaluate the quad GM-PHD filter as a typical example and compare it to four independent GM-PHD filters, indicating that our approach shows better performance determined using cardinality, OSPA metric and discrimination rate among the different target types. Even though, we show the simulation analysis for $N = 4$, in principle the methodology can be applied to $N$ types of targets where $N$ is a variable in which the number of possible confusions may rise as $N(N-1)$. For instance, after experimenting the dual GM-PHD filter ($N = 2$) and the tri-GM-PHD filter ($N = 3$) by simulation and making sure that they show similar behaviour as for $N=4$, we applied the tri-GM-PHD filter for visual tracking applications in~\cite{BaiWal17}. In case there is no target confusion, the N-type GM-PHD filter performs similar to $N$ independent GM-PHD filters. On the other hand, if each target is regarded as a type, for instance each of the four target types in this given example has only one target, the N-type GM-PHD filter is used as a labeler of each target. This means it discriminates each target from frame to frame (from one time step to another) rather than simply degrading to the standard GM-PHD filter(s).

\section{Appendix A} \label{sec:AppendixA}

PHD filter for multi-target tracking of a single target type was proposed by Mahler in~\cite{Mah03} where its derivation has been given. We have derived novel extensions for the updated PHDs of a N-type PHD filter from PGFLs of each target type handling confusions among target types starting from the standard proved predicted PHDs in section~\ref{Subsec:PGFL}. In this case we assume the standard predicted PHD can be applied to each target type as influence of the measurement confusions among the target types we are solving occurs at the update stage of the PHD recursion. To make this paper self-sufficient, we derive the predicted PHDs for each target type $i \in \{1,...,N\}$ starting from the PGFLs of the predicted processes. We use the product rule and chain rule for functional derivatives to find PHD prediction equation for target type $i \in \{1,...,N\}$ (differentiate and set $h=1$) as follows:

\begin{equation}
    G_{i,k|k-1} (h) = G_{i,\gamma} (h) G_{i,k-1}(G_{i,S}(h|.)),
\label{eqn:PGFLp1}
\end{equation}
\noindent where (taking $\varphi_x = \delta_x$)
\begin{equation}
    \frac{\delta}{\delta\varphi_x} G_{i,\gamma} (h)\bigg|_{h=1}  = \gamma_{i,k}(x).
\label{eqn:PGFLp1b}
\end{equation}
\noindent
\begin{equation}
    \frac{\delta}{\delta\varphi_x} G_{i,k-1} (h)\bigg|_{h=1}  = \mathcal{D}_{i,k-1}(x).
\label{eqn:PGFLp1pr}
\end{equation}
\noindent and
\begin{equation}
    G_{i,S} (h|x) = 1 - p_{i,S,k}(x) + p_{i,S,k}(x)\int h(y) y_{i,k|k-1}(y|x)dy.
\label{eqn:PGFLp1s}
\end{equation}
\noindent the aim is to compute the PHD of the predicted process $D_{i,k|k-1}(x)$ for target type $i \in \{1,...,N\}$.

Deriving Eq.~\ref{eqn:PGFLp1} in $\varphi_x = \delta_x$ and using the product rules yields:

\begin{widetext}
\begin{equation}
\begin{array} {lll}
    \frac{\delta}{\delta\varphi_x} G_{i,k|k-1} (h)\bigg|_{h=1} =& \underbrace{\frac{\delta}{\delta\varphi_x} G_{i,\gamma} (h)\bigg|_{h=1}}_{=\gamma_{i,k}(x)} \underbrace{G_{i,k-1}(G_{i,S}(h|.))\big|_{h=1}}_{=1} + \underbrace{G_{i,\gamma} (h)\big|_{h=1}}_{=1} \underbrace{\frac{\delta}{\delta\varphi_x} G_{i,k-1}(G_{i,S}(h|.))\bigg|_{h=1}}_{=A}
\end{array}
\label{eqn:PGFLp1productRules}
\end{equation}
\end{widetext}
\noindent

Now, using the chain rule $A$ reads:
\begin{equation}
    A = \frac{\delta G_{i,k-1}}{\delta\big(\frac{\delta G_{i,S}(h|.)}{\delta\varphi_x}\big)} (G_{i,S}(h|.))\bigg |_{h=1}
\label{eqn:PGFLp1productRulesA}
\end{equation}
\noindent where $\frac{\delta G_{i,S}(h|.)}{\delta\varphi_x}$ is found by deriving Eq.~\ref{eqn:PGFLp1s} and using the linearity of $h \rightarrow \int h(y)y_{i,k|k-1}(y|.)dy$ as follows:

\begin{equation}
\begin{array} {lll}
    \frac{\delta G_{i,S}(h|.)}{\delta\varphi_x} =& \underbrace{\frac{\delta}{\delta\varphi_x} (1 - p_{i,S,k}(.))}_{=0} + \\& p_{i,S,k}(.) \underbrace{\frac{\delta}{\delta\varphi_x}\int h(y) y_{i,k|k-1}(y|.)dy}_{=y_{i,k|k-1}(x|.)}.
\end{array}
\label{eqn:PGFLp1sD}
\end{equation}
\noindent

Let us write $g(.) = p_{i,S,k}(.) y_{i,k|k-1}(x|.)$ for simplicity's sake. Now all we need to do is to compute $\frac{\delta G_{i,k-1}}{\delta g(.)}(G_{i,S}(h|.))\bigg|_{h=1}$. Using the Janossy densities we can write:

\begin{equation}
    G_{i,k-1} (h) = \sum_{n=0}^{\infty} \frac{1}{n\,!} \int_{\mathcal{X}^n} \bigg(\prod_{l=1}^{n}h(x_l)\bigg)j_{i,k-1}^{(n)} (x_1,...,x_n) dx_1...dx_n.
\label{eqn:PGFLp1prJ}
\end{equation}
\noindent

Deriving $G_{i,k-1}$ in $g(.)$ gives:

\begin{widetext}
\begin{equation}
\begin{array} {lll}
    \frac{\delta G_{i,k-1}}{\delta g(.)} (h) \bigg|_{h=1} &= \sum_{n=1}^{\infty} \frac{1}{n\,!} \int_{\mathcal{X}^n} \frac{\delta}{\delta g(.)} \bigg(\prod_{l=1}^{n}h(x_l)\bigg)\bigg|_{h=1}j_{i,k-1}^{(n)} (x_1,...,x_n) dx_1...dx_n, \\
       &= \sum_{n=1}^{\infty} \frac{1}{n\,!} \int_{\mathcal{X}^n} \sum_{l=1}^{n} \bigg(\frac{\delta}{\delta g(.)}h(x_l)\prod_{j\neq l}h(x_j)\bigg)\bigg|_{h=1}j_{i,k-1}^{(n)} (x_1,...,x_n) dx_1...dx_n.
\end{array}
\label{eqn:PGFLp1prJD}
\end{equation}
\end{widetext}
\noindent

Using the definition of functional derivatives with $F[h] = h(x_l)$ gives:

\begin{widetext}
\begin{equation}
\begin{array} {lll}
    \frac{\delta G_{i,k-1}}{\delta g(.)} (h) \bigg|_{h=1} &= \sum_{n=1}^{\infty} \frac{1}{n\,!} \int_{\mathcal{X}^n}  \bigg(\sum_{l=1}^{n} g(x_l)\bigg)j_{i,k-1}^{(n)} (x_1,...,x_n) dx_1...dx_n, \\
       &= \sum_{n=1}^{\infty} \frac{1}{n\,!} \sum_{l=1}^{n} \int_{\mathcal{X}^n}  g(x_l)j_{i,k-1}^{(n)} (x_1,...,x_n) dx_1...dx_n.
\end{array}
\label{eqn:PGFLp1prJD2}
\end{equation}
\end{widetext}
\noindent

That is, since the Janossy densities are symmetric functions:

\begin{widetext}
\begin{equation}
\begin{array} {lll}
    \frac{\delta G_{i,k-1}}{\delta g(.)} (h) \bigg|_{h=1} &= \sum_{n=1}^{\infty} \frac{1}{(n-1)\,!} \int_{\mathcal{X}^n}  g(x_1)j_{i,k-1}^{(n)} (x_1,...,x_n) dx_1...dx_n, \\
     &=  \int_{\mathcal{X}} g(x_1) \bigg(\sum_{n=1}^{\infty} \frac{1}{(n-1)\,!} \int_{\mathcal{X}^{n-1}} j_{i,k-1}^{(n)} (x_1,...,x_n) dx_2...dx_n \bigg)dx_1,\\
     &= \int_{\mathcal{X}} g(x) \bigg(\sum_{n=0}^{\infty} \frac{1}{n\,!} \int_{\mathcal{X}^{n}} j_{i,k-1}^{(n+1)} (x,x_1,...,x_n) dx_1...dx_n \bigg)dx.
\end{array}
\label{eqn:PGFLp1prJD3}
\end{equation}
\end{widetext}
\noindent

Or, using the characterization of the PHD $\mathcal{D}_{i,k-1}(x)$ as the first-moment density of the process:

\begin{equation}
    \frac{\delta G_{i,k-1}}{\delta g(.)} (h) \bigg|_{h=1} = \int_{\mathcal{X}}g(x)\mathcal{D}_{i,k-1}(x)dx.
\label{eqn:PGFLp1prJD4}
\end{equation}
\noindent

Combining Eq.~\ref{eqn:PGFLp1productRules}, Eq.~\ref{eqn:PGFLp1productRulesA}, Eq.~\ref{eqn:PGFLp1sD} and Eq.~\ref{eqn:PGFLp1prJD4} yields the result:

\begin{widetext}
\begin{equation}
    \mathcal{D}_{i,k|k-1}(x) = \frac{\delta G_{i,k|k-1}}{\delta \varphi_x} (h) \bigg|_{h=1} = \gamma_{i,k}(x) + \int_{\mathcal{X}} p_{i,S,k}(\zeta)y_{i,k|k-1}(x|\zeta)\mathcal{D}_{i,k-1}(\zeta)d\zeta.
\label{eqn:PGFLp1Pfnl}
\end{equation}
\end{widetext}
\noindent

% or


\section*{Acknowledgment}

We would like to acknowledge the support of the Engineering and Physical Sciences Research Council (EPSRC), grant references EP/K009931 and a James Watt Scholarship. We would also like to thank Dr. Daniel Clark for sharing his expertise and understanding of RFS methodology.

\ifCLASSOPTIONcaptionsoff
  \newpage
\fi

\bibliographystyle{IEEEtran}
%\bibliography{strings,refs}
\bibliography{egbib}


%\vfill


\end{document}